\documentclass[12pt]{iopart}

\usepackage{scalerel}[2014/03/10]
\usepackage{stackengine}
\usepackage{amsmath}
\usepackage{hyperref}
\usepackage{subcaption}
\usepackage{amsfonts}
\hypersetup{
    colorlinks=true,
    linkcolor=blue,
    filecolor=magenta,      
    urlcolor=cyan,
    pdftitle={Overleaf Example},
    pdfpagemode=FullScreen,
  }

\usepackage{cite}

\usepackage{cancel}
\usepackage{bm}
\usepackage{physics}

\newcommand{\av}[1]{\left\langle #1 \right\rangle}

\newcommand{\sgn}{\operatorname{sgn}}

\newcommand{\ext}{\text{ext}}

\newcommand{\sts}{\text{s}}

\newcommand{\calX}{\mathcal{X}}

\newcommand{\eq}{\text{eq}}
\newcommand{\eff}{\text{eff}}
\newcommand{\modulo}{\text{mod}}
\newcommand{\ub}{\modulo(x+\Delta x,L)}
\newcommand{\lb}{\modulo(x-\Delta x,L)}

\newcommand{\Pb}{P(x,t_k^-)}

\newcommand{\Pperxb}{\tilde{P}(x,t_k^-)}

\newcommand{\Pcb}{P(x,c,t_k^-)}
\newcommand{\Pca}{P(x,c,t_k^+)}
\newcommand{\Ppera}{\tilde{P}(x,c,t_k^+)}
\newcommand{\Pperb}{\tilde{P}(x,c,t_k^-)}

\newcommand{\PsXC}{P_{\sts}^{(c)}(x)}
\newcommand{\PsXprimeC}{P_{\sts}^{(c)}(x')}

\newcommand{\PsXone}{P_{\sts}^{(1)}(x)}
\newcommand{\PsXzero}{P_{\sts}^{(0)}(x)}

\newcommand{\J}{J(x,c,t)}

\newcommand{\Jper}{\tilde{J}(x,c,t)}

\newcommand{\Pper}{\tilde{P}(x,c,t)}

\newcommand{\Pc}{P(x,c,t)}

\newcommand{\dH}{\dot{H}(t)}

\newcommand{\period}{\text{per}}

\newcommand{\calL}{\mathcal{L}}

\begin{document}

\title[Markovian description of feedback-controlled systems]{Markovian description of a wide class of feedback-controlled systems: Application to the feedback flashing ratchet}

\author{Natalia Ruiz-Pino and  Antonio Prados}
\address{Física Teórica, Universidad de Sevilla, Apartado de Correos
  1065, E-41080 Sevilla, Spain}
\ead{nruiz1@us.es, prados@us.es}
\vspace{10pt}
\begin{indented}
\item[]\today
\end{indented}

\begin{abstract}
In feedback-controlled systems, an external agent---the feedback controller---measures the state of the system and modifies its subsequent dynamics depending on the outcome of the measurement. In this paper, we build a Markovian description for the joint stochastic process that comprises both the system and the controller variables. This Markovian description is valid for a wide class of feedback-controlled systems, allowing for the inclusion of errors in the measurement. The general framework is motivated and illustrated with the paradigmatic example of the feedback flashing ratchet.
\end{abstract}

%
%
%
\maketitle
%
%

\section{Introduction}\label{sec:intro}

The dynamics of a physical system typically depends on the system variables plus some additional quantities, the time dependence of which can often be externally controlled. For instance, the evolution of the position of an overdamped Brownian particle in an optical harmonic trap depends both on the stiffness of the trap and the temperature of the bath. The stiffness of the trap can be tuned by varying the intensity of the laser beam, and the temperature of the bath can be effectively increased by using charged particles and introducing an additional random electric field---both procedures are now standard in the manipulation of colloidal particles~\cite{martinez_colloidal_2017,ciliberto_experiments_2017}. 

By engineering the time evolution of these additional quantities---the so-called control functions, one usually aims at controlling the time evolution of the system. The simplest possibility is to impose a certain time protocol for these control functions, independently of the instantaneous state of the system: this is what is known as \textit{open-loop control}. For example, this is the case of the shortcuts to adiabaticity in quantum systems~\cite{guery-odelin_shortcuts_2019} and the analogous swift state-to-state transformations in classical and stochastic systems~\cite{guery-odelin_driving_2023}, in which open-loop control is employed to accelerate the system dynamics.

A more sophisticated approach is that of \textit{feedback control}, in which the time protocol of the control function depends on the state of the system~\cite{bechhoefer_feedback_2005,sagawa_thermodynamics_2012,sagawa_nonequilibrium_2012,annby-andersson_quantum_2022}. Ideally, the feedback controller measures without any error the instantaneous state of the system, without any time delay, and thus acts on the system dynamics without any uncertainty. Feedback control allows for improving the performance of the system, e.g. increasing the flux in Brownian ratchets~\cite{cao_feedback_2004,cao_information_2009,feito_rocking_2009}. Still, this improvement of the performance has a thermodynamic cost, since the information gathered by the controller when measuring the system state has to be incorporated to the thermodynamic balance~\cite{cao_thermodynamics_2009,sagawa_minimal_2009,abreu_thermodynamics_2012,sagawa_thermodynamics_2012,sagawa_nonequilibrium_2012,deffner_information_2013,van_vu_uncertainty_2020,ruiz-pino_information_2023}.

The action of the feedback controller makes the system dynamics non-Markovian, since the instantaneous values of the system variables do not suffice to univocally determine its subsequent evolution: this evolution also depends on the values of the variables characterising the state of the controller, which in turn depend on the value of the system variables at the last time that the control measured the system state. Note that this is so even for feedback controllers without any time delay, which are sometimes called Markovian feedback controllers. The non-Markovianity of the system dynamics complicates its mathematical description. For example, the full chain of control actions must be taken into account in the thermodynamic balance~\cite{ruiz-pino_information_2023}.

Despite the above, in this paper we build a Markovian description for a wide class of feedback controlled systems, by considering the joint process of the system and controller variables. The spirit of our approach is thus similar to recent works on feedback-controlled quantum systems~\cite{annby-andersson_quantum_2022}, {and also in stochastic systems with continuous measurement-feedback~\cite{horowitz_thermodynamics_2014}. {{ For periodic measurement feedback, a two-level system controlled in a discrete feedback loop has also been modelled in terms of stochastic Markov processes~\cite{um_total_2015}.}} For the sake of simplicity, we present the framework for an overdamped particle in the one-dimensional case; the extension to $d$ dimensions and/or underdamped dynamics is straightforward.}

We consider an overdamped particle with position $x$ moving in a fluctuating potential $V(x,c)$, the shape of which depends on an additional variable $c$.  The feedback controller measures the instantaneous state of the system just before certain given times $t_k$, $x_k\equiv x(t_k^-)$, $k=1,2,\ldots$, and the controller variable changes to the value $c_k$ following a certain probability distribution $\Theta(c_k|x_k)$---the flashing protocol. Then, the potential acting on the particle in the time interval between the $k$-th and $(k+1)$-th measurements, $I_k=(t_k,t_{k+1})$, is given by $V(x,c_k)$. The above framework entails that there is no time delay, since the potential $V(x,c_k)$ is already felt by the particle for $t=t_k^+$. Still, our framework allows for the introduction of error, or uncertainty, in the measurement---a possibility that has been considered in the literature~\cite{cao_information_2009,dinis_extracting_2020}.

Our work is motivated by, and later applied to, a feedback-controlled flashing ratchet. Brownian ratchets rectify Brownian motion by creating flux in a given direction, by breaking the left-right symmetry~\cite{astumian_brownian_2002,reimann_brownian_2002,reimann_introduction_2002,parrondo_energetics_2002}. In addition to their theoretical importance, they are also relevant in biosystems, as models for molecular motors~\cite{astumian_mechanochemical_1996,serreli_molecular_2007,kay_synthetic_2007,erbas-cakmak_artificial_2015,kassem_artificial_2017,grelier_unlocking_2024}. In the open-loop (no feedback) case, the breaking of spatial symmetry can be brought about by the consideration of an asymmetric potential, which is periodically switched on and off~\cite{ajdari_mouvement_1992,magnasco_forced_1993,astumian_fluctuation_1994,rowchowdhury_maximizing_2012}. Instead of switching on and off the potential, the breaking of spatial symmetry can also be brought about by the variation of the amplitude and period of a force acting on the particle, as in the deterministic ratchet of Ref.~\cite{cubero_overdamped_2006}. Several studies have been carried out to elucidate how the degree of symmetry breaking impacts on the optimisation of transport in both deterministic~\cite{chacon_criticality-induced_2010} and Brownian~\cite{martinez_ratchet_2013} ratchets.

In general, closed-loop or feedback control makes it possible to increase the performance of ratchets and other stochastic systems~\cite{touchette_information-theoretic_2000,touchette_information-theoretic_2004,cao_feedback_2004,cao_information_2009,ruiz-pino_information_2023, abreu_thermodynamics_2012,lucero_maximal_2021}.  Therein, spatial symmetry can be broken even for symmetric potentials, the feedback mechanism breaks it by switching on or off the potential depending on the sign of the force at the particle's measured position. In the specific example of the feedback flashing ratchet that we employ to illustrate the general framework developed in this work, we consider a symmetric V-shaped potential and a flashing protocol $\Theta(c|x)$ that incorporates error in the measurement in a realistic way. Yet, analytical predictions can be obtained for the time evolution of the system, which compare excellently with numerical simulations of the dynamics.

Our paper is organised as follows. Section~\ref{sec:motivation-feedback-ratchet} motivates our work by considering the feedback flashing ratchet, neatly showing that $X$ is not Markovian but the joint process $(X,C)$ is Markovian. The general framework is presented in Sec.~\ref{sec:general-frame}, in which a differential Chapman-Kolmogorov equation for the joint process is derived. In Sec.~\ref{sec:H-theorem}, we prove an $H$-theorem for the joint process $(X,C)$, which entails the emergence of a long-time regime independent of the initial conditions. Section~\ref{sec:coarse-grain-Markov-x} considers a coarse-grained timescale, in which the position $X$ becomes Markovian. Therein, we also consider some relevant limit cases and discuss the connection between the fine and coarse timescales. The general framework developed is illustrated in Sec.~\ref{sec:flashing-ratchet} with the particular case of the feedback flashing ratchet. Numerical simulations and theoretical results are compared, finding an excellent agreement between them. The main results of the paper are discussed in Sec.~\ref{sec:conclusions}, together with perspectives for future work. The Appendices deal with some technical details that are omitted in the main text.

\section{Motivation: feedback flashing ratchet}\label{sec:motivation-feedback-ratchet}

Our work is motivated by the so-called \textit{feedback flashing ratchet}~\cite{cao_feedback_2004,dinis_closed-loop_2005,lopez_realization_2008,feito_rocking_2009,jarillo_efficiency_2016,kim_deep_2021,ruiz-pino_information_2023}. This system comprises an overdamped Brownian particle immersed
in a fluid at equilibrium with temperature $T$. The friction
coefficient is $\gamma$, which is related to the diffusion
coefficient $D$ by the Einstein fluctuation-dissipation relation
$D=k_{B}T/\gamma$. The Brownian particle moves on a line under the
action of a conservative force, deriving from a periodic
potential $V(x)$, and a non-conservative force.  The motion of the
particle can be studied either in the whole real line
$-\infty<x<+\infty$, or in a finite interval $-L\leq x\leq L$---in the
latter case with periodic boundary conditions, note that the period of the potential is $2L$, see~\ref{app:periodic-bc}.

The potential is switched on or off by an external feedback controller, which measures the instantaneous position of the particle at regular time intervals $\Delta t_{m}$ and makes the switching depending thereon. More specifically, the feedback controller measures the position of the particle $x_k^-\equiv x(t_{k}^{-})$ just before $t_{k}=k\Delta t_{m}$, $k=1,2,\ldots$. These measurements naturally divide the timeline into intervals $I_{k}=(t_{k},t_{k+1})$ of length $\Delta t_{m}$, so that $t_{k}^{-}\in I_{k-1}$. After the measurement, the potential is on (off) in the next time interval $I_{k}$ if the force $-V'(x_k^-)>0$ ($-V'(x_k^-)<0$). This decision on the switching creates the ratchet effect, i.e. a directed movement to the right\cite{reimann_brownian_2002,reimann_introduction_2002,astumian_brownian_2002}. This may be used to exert work against the external non-conservative force, which typically opposes the average direction of motion.

\begin{figure}
    \centering
    \includegraphics[width=0.9\linewidth]{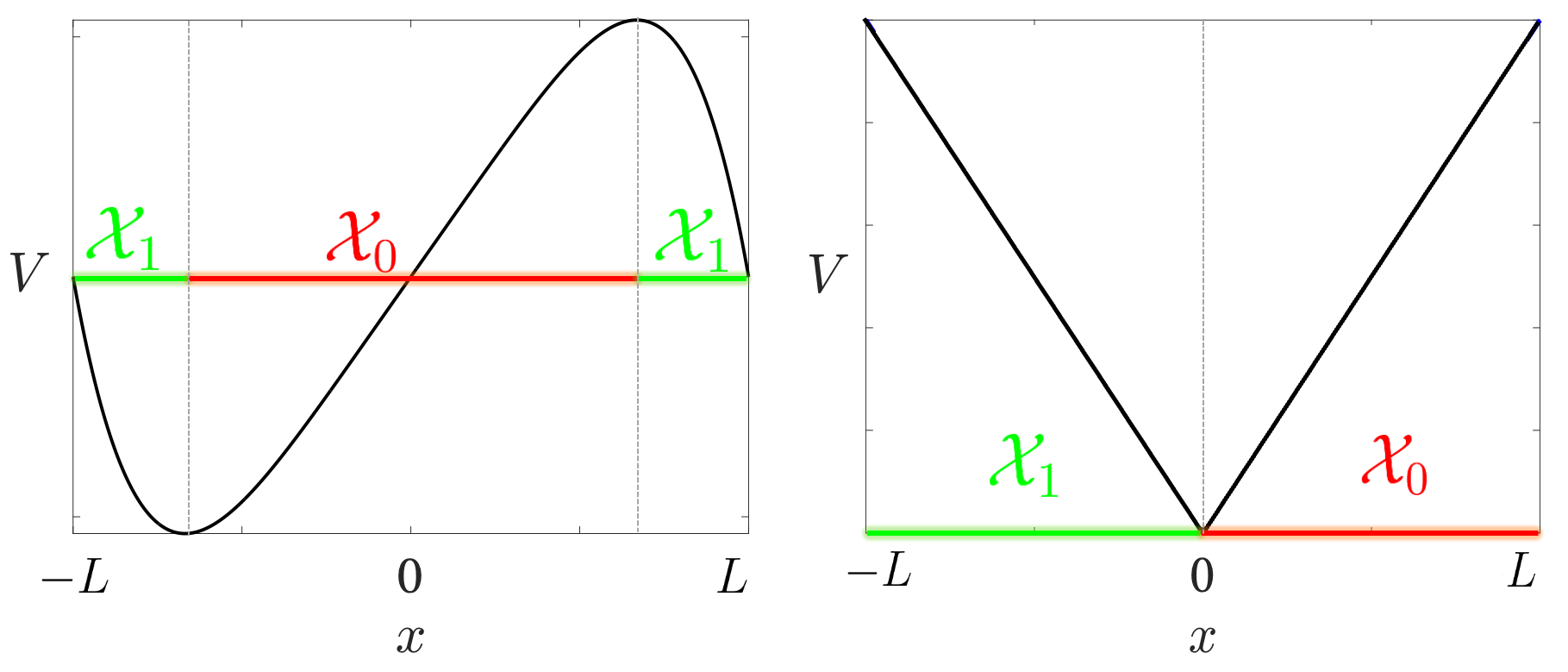}
    \caption{Intervals $\calX_0$ and $\calX_1$ for the flashing protocol for different potentials. The interval $\calX_1$ corresponds to those positions for which the force is positive, $-\partial_x V>0$, and  $\calX_0$ for those for which the force is negative, $-\partial_x V<0$. The external controller measures the particle position and the potential is switched on (off) afterwards, i.e. the potential felt by the particle is $c\, V(x)$, with $c=1$ ($c=0$),  if the position of the particle belongs to $\calX_1$ ($\calX_0$).  }
    \label{fig:V-shaped-potential}
\end{figure}

The state of the controller can be characterised by a dichotomic
variable $C$, with possible values $c=(0,1)$, such that the potential
is on (off) for $c=1$ ($c=0$). In other words, the potential felt by
the particle is $V(x,c)=c\,V(x)$. We may define two sets $\calX_{1}$
and $\calX_{0}$, such that:
\begin{align}\label{eq:X0-X1}
  x\in\calX_{1} \text{ if } -V'(x)>0, \quad
  x\in\calX_{0} \text{ if } -V'(x)<0.
\end{align}
It is clear that $\calX_{0} \cup \calX_{1}=[x_{L},x_{R}]$,
where $[x_{L},x_{R}]$ is the complete interval in which the movement
is studied. The specific case of a V-shaped potential, which we later use in Sec.~\ref{sec:flashing-ratchet} for the numerical check of our theoretical framework, is illustrated in Fig.~\ref{fig:V-shaped-potential}. 

The above picture entails that, in
the time interval $I_k$ between control updates, the Brownian particle moves under the potential $V(x,c_{k})=c_{k}V(x)$, where
$x(t_{k}^{-})\in \calX_{c_{k}}$. Therefore, it obeys the Langevin equation
\begin{equation}
\gamma\dot{x} = -c_k\, V'(x)-F_{\ext}+\xi(t), \quad t\in I_{k},
\label{eq:Langevin}
\end{equation}
 where $\xi(t)$ is the Gaussian white noise:
\begin{equation}
  \langle \xi(t)\rangle=0, \quad \langle \xi(t)\xi(t')\rangle =
  2 k_B T \gamma \, \delta(t-t'),
\label{eq:white-noise}
\end{equation}
with $k_B$ being the Boltzmann constant. We recall that the
non-conservative force faces the opposite direction of the average
motion of the particle to the right. This is the reason why we write
$-F_{\ext}$ in Eq.~\eqref{eq:Langevin}, with $F_{\ext}>0$.

Despite obeying the Langevin equation \eqref{eq:Langevin} between
control updates, the stochastic process $X$ is not Markovian: its time
evolution has memory of earlier times through the value of the control
variable $c_{k}$, which depends on $x(t_{k}^{-})$. The value
of $x(t_{k}^{-})$ depends in turn on the previous value of the control
$c_{k-1}$, and so on. Therefore, in order to analyse the thermodynamic
balance of the system, the whole chain of control actions has been
typically considered---which complicates the description of the
system~\cite{ruiz-pino_information_2023}.

Here, we argue a fact that seems to have been overlooked in previous
analysis of the feedback flashing ratchet: although the process $X$
alone is not Markovian, the joint process $(X,C)$ is
Markovian. Moreover, this Markovian description for the joint process
is valid for a wide class of feedback controlled systems, not only for
error-free measurements but also for systems where the action of the
controller allows for errors or has some degree of randomness.  This
is the reason why, in the following section, we develop a general
theoretical framework, which is afterwards applied to
the feedback flashing ratchet.

\section{General framework}\label{sec:general-frame}

In this section, we put forward our general formalism for describing a
wide class of feedback-controlled systems. This formalism includes, as
a particular case, the feedback flashing ratchet with error-free
measurements described in the previous section, but also allows for
the inclusion of errors in the measurement~\cite{cao_information_2009,dinis_extracting_2020}.
In addition, more complex situations in which the variable characterising the
controller is not dichotomic fit into the general framework developed below, i.e. we can incorporate feedback to situations in which the potential fluctuates between more than two states, as in Refs.~\cite{bier_biasing_1996,li_escape_2006,mankin_thermally_2008}.

Let us consider an overdamped Brownian particle, immersed in a fluid
at equilibrium with temperature $T$. The friction coefficient is
$\gamma$, which is related to the diffusion coefficient $D$ by the
Einstein fluctuation-dissipation relation $D=k_{B}T/\gamma$. The
motion of the particle takes place in a certain interval
$[x_{L},x_{R}]$---which may be the whole line, $x_{L}=-\infty$,
$x_{R}=+\infty$, or a finite interval of size $2L$, e.g. $x_{L}=-L$,
$x_{R}=L$.

The particle moves under the action of a conservative force, deriving
from a fluctuating potential, and a non-conservative force. The
fluctuation of the potential is accounted for by a variable $c$,
$V(x,c)\ne V(x,c')$ if $c\ne c'$. This fluctuation is controlled by an
external feedback mechanism, which uses information about the position
of the particle $x$ to decide the value of $c$---which thus
characterises the state of the controller.  When the controller
variable equals $c$, the total force acting on the particle is
\begin{equation}
  -\partial_{x}\tilde{V}(x,c)=-\partial_{x}V(x,c)-F_{\ext} .
\end{equation}

Here, we assume that the value of the control variable $c$ is updated
at regular times $t_{k}=k\Delta t_{m}$, $k=0,1,2,\ldots$. This
naturally divides the timeline into intervals $I_{k}=(t_{k},t_{k+1})$,
over which the control variable $c$ takes a constant value
$c_{k}=c(t_{k}^{+})$. Over each trajectory of the stochastic dynamics,
the controller measures the position of the particle just before the
update, $x_{k}^{-}\equiv x(t_{k}^{-})$, and the potential takes the
shape $V(x,c_{k})$ with a certain probability
$\Theta(c_{k}|x_{k}^{-})$ throughout the next time interval
$I_{k}$. Normalisation entails that
\begin{equation}\label{eq:thetac-norm}
  \sum_{c}\Theta(c|x)=1, \quad \forall x.
\end{equation}

In the typical example of the feedback flashing ratchet, described in the previous section, the potential is switched on or off without any error ~\cite{cao_feedback_2004,cao_thermodynamics_2009, ruiz-pino_information_2023}, depending on the sign of the conservative force---see Eq.~\eqref{eq:X0-X1}. In that case, we have
\begin{align}
  \Theta(c|x)=\begin{cases}
    1, & \text{if } x\in\calX_{c}, \\
    0, & \text{if } x\not\in\calX_{c}.
    \label{eq:theta-c-perfect}
  \end{cases}
\end{align}
This expression also applies to any error-free feedback mechanism, which measures the position of the
particle $x_{k}^{-}$ and always chooses one of the potentials
$V(x,c)$ during the subsequent time interval $I_{k}$, depending on the interval $\calX_{c}$ to which $x_{k}^{-}$
belongs. 

\begin{figure}
    \centering  \includegraphics[width=0.75\linewidth]{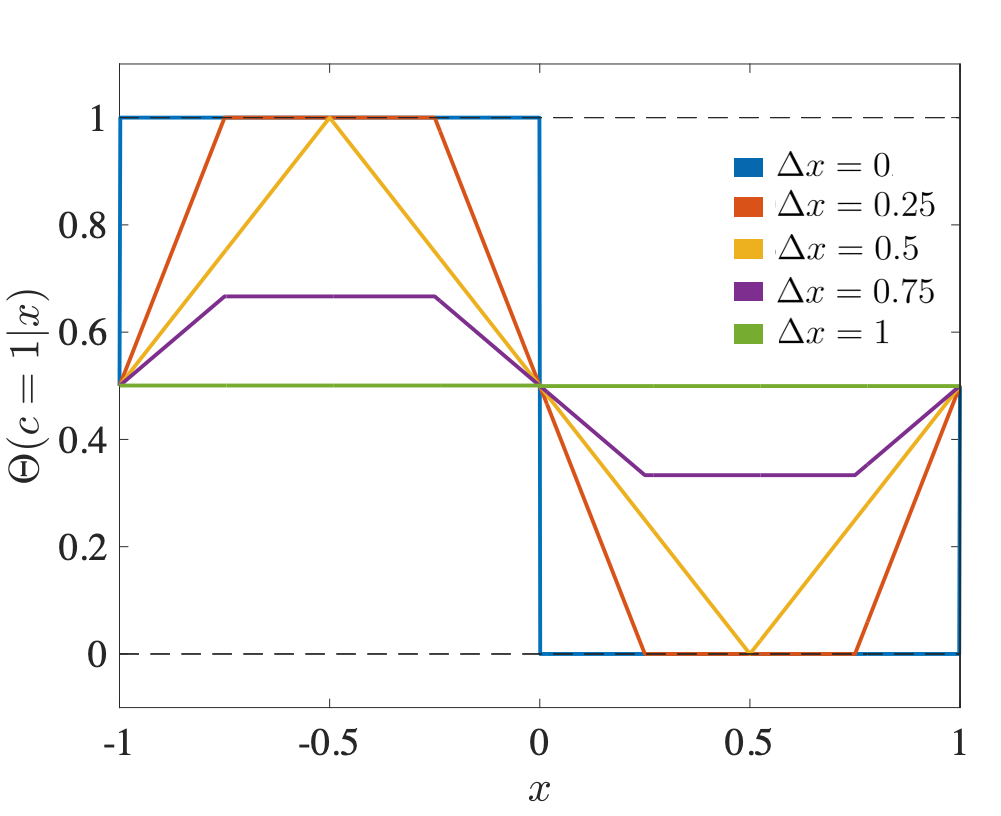}
    \caption{Feedback control protocols  for different values of the uncertainty in the measurement. When $\Delta x=0$, the feedback control protocol corresponds to an error-free one (blue line). As the uncertainty $\Delta x$ increases, the protocol continuously converges  to the open-loop protocol, which is attained for $\Delta x=L$ (green line). At this point, the probability of switching on the potential becomes independent of the position and equals the probability of switching it off, i.e. it is equal to $1/2$.}
    \label{fig:protocols}
\end{figure}

More realistically, the particle position is measured with a certain uncertainty $\Delta x$, i.e. we know that it lies on a segment $(x-\Delta x, x+\Delta x)$. With this information, the controller chooses the potential $V(x,c)$ with a probability proportional to the length of the intersection between $(x-\Delta x, x+\Delta x)$ and $\calX_{c}$: \begin{align}
  \Theta(c|x)=\frac{1}{2\Delta x}\text{length\;} \{\calX_{c} \cap (x-\Delta x, x+ \Delta x)\}
  \label{eq:theta-c-imperfect}
\end{align}
For $\Delta x=0$, Eq.~\eqref{eq:theta-c-imperfect} reduces to the error-free case in Eq.~\eqref{eq:theta-c-perfect}. On the other hand, for the periodic case in which the particle moves in a finite interval $[-L,L]$, the case $\Delta x = L$ corresponds to an open-loop protocol: the control switches the potential with a probability proportional to the length of the interval $\calX_c$, independently of the particle position.  Equation~\eqref{eq:theta-c-imperfect} is illustrated in Fig.~\ref{fig:protocols}, for the case of the simple V-shaped potential depicted in Fig.~\ref{fig:V-shaped-potential}. 

We would like to remark that the general framework developed in the next section is valid for an arbitrary choice of $\Theta(c|x)$, as long as Eq.~\eqref{eq:thetac-norm} holds. This includes non-dichotomic feedback controllers, whose variable $c$ takes several values corresponding to different states of the potential. It also allows for open-loop control, which occurs when $\Theta(c|x)$ is independent of $x$, i.e. $\Theta(c|\cancel{x})=\Theta(c)$.

\subsection{Markovian description for the joint position-control
  stochastic process}\label{sec:dynamic-eq}

Now we write down an evolution equation for the joint probability
distribution $\Pc$. In the time intervals $I_{k}=(t_{k},t_{k+1})$, i.e. between
control updates, the control is ``static'' and thus $\Pc$ obeys
the Fokker-Planck equation with potential $V(x,c)$, i.e.
\begin{equation}\label{eq:FP-Jk}
     \partial_t \Pc=-\partial_x \J, \quad t\in I_{k},
\end{equation}
where the probability flux is defined from
\begin{align}
  -\gamma \J\equiv & \partial_{x}\tilde{V}(x,c) \, \Pc
                           + k_B T \, \partial_x \Pc .
                           \label{eq:J-def}
\end{align}

On top of the dynamics over the time interval $I_{k}$, we have the
update of the control values at times $t_{k}$.  After measuring the
position of the particle $x_{k}^{-}$ at time $t_{k}^{-}$, we set the
value $c$ with probability $\Theta(c|x_{k}^{-})$. We build $P(x,c,t_{k}^{+})$ in
terms of $P(x,c,t_{k}^{-})$ by thinking of the joint probability
$\Pc$ as the density of walkers at $(x,t)$ with the control
variable equal to $c$. Since the decision on the value of $c$ at
$t_{k}^{+}$ only depends on the value of the position at $t_{k}^{-}$,
we have that 
 \begin{align}
    P(x,c,t_{k}^{+})=\Theta(c|x) P(x,t_{k}^{-}),
    \label{eq:P-tk-+}
 \end{align}
 where
 \begin{equation}\label{eq:Px-marginal}
   P(x,t)=\sum_{c}\Pc
 \end{equation}
 is the marginal distribution of the
 particle position. Note that $P(x,c,t_{k}^{+})$ is normalised as a
 consequence of $P(x,t_{k}^{-})$ and $\Theta(c|x)$ being
 normalised.

The meaning of Eq.~\eqref{eq:P-tk-+} is clear: at position $x$
 and time $t_{k}^{+}$, the number of walkers moving under the
 potential $V(x,c)$ equals the number of walkers at position $x$ at
 time $t_{k}^{-}$ multiplied by the probability $\Theta(c|x)$. Note
 that the marginal distribution $P(x,t)$ is continuous at time
 $t_{k}$,
\begin{equation}
  \label{eq:P-tk-+-marginal-continuity}
  P(x,t_{k}^{+})=P(x,t_{k}^{-})
  \sum_{c}\Theta(c|x) = P(x,t_{k}^{-}),
\end{equation}
which is logical: the density of walkers at $x$ is unaffected by the
update of the control value. Henceforth, we can thus write
$P(x,t_{k})$ without any ambiguity. The conditional probability of the
control values just after the control update is precisely
$\Theta(c|x)$,
\begin{equation}
  \label{eq:P-cond-C-on-x}
  P(c|x,t_{k}^{+})=\frac{P(x,c,t_{k}^{+})}{P(x,t_{k}^{+})}=\Theta(c|x). 
\end{equation}
Thus, for the error-free measurement~\eqref{eq:theta-c-perfect}, the
value of the control at $t_{k}^{+}$ is enslaved to the value of the
position. %

At the control updates, between $t=t_{k}^{-}$ and $t=t_{k}^{+}$, the
joint probability distribution evolves according to
\begin{align}
 P(x,c,t_{k}^{+})-P(x,c,t_{k}^{-})= \Theta(c|x)\sum_{c'\ne
c}P(x,c',t_{k}^{-})-\left[1-\Theta(c|x)\right]P(x,c,t_{k}^{-}),
\end{align}
where we have made use of Eqs.~\eqref{eq:P-tk-+} and
\eqref{eq:Px-marginal}. The first term on the rhs is the typical gain
term, due to transitions from $c'$ to $c$, weighted with probability
$\Theta(c|x)$, whereas the second term is the typical loss term,
weighted with probability $1-\Theta(c|x)$. Therefore, we can write the following evolution equation for the joint
probability $\Pc$:
\begin{align}
  \partial_{t}\Pc=& -\partial_{x}\J \nonumber \\
  & +\sum_{k}\delta(t-t_{k}^{-})
                                      \Big[\Theta(c|x)\sum_{c'\ne
c}P(x,c',t)-\left[1-\Theta(c|x)\right]\Pc \Big].
    \label{eq:diff-CK}
\end{align}

Equation~\eqref{eq:diff-CK} has the form of the differential Chapman-Kolmogorov equation (DCKE)---see section 3.4 of Ref.~\cite{gardiner_stochastic_2009}---and it is thus the mathematical expression of the fact that $(X,C)$ is a Markov
process. On the one hand, the first term on the rhs corresponds to the Fokker-Planck evolution between control updates. On the other hand, the second term on the rhs is a master equation term  with time-dependent transition rates,
which are time-periodic with period $\Delta t_{m}$. Therefore, one
expects $\Pc$ to tend to a time-periodic state with the same
period in the long-time limit, when initial conditions are forgotten. This physical expectation is confirmed in Sec.~\ref{sec:H-theorem} by our deriving an $H$-theorem for Eq.~\eqref{eq:diff-CK}. 

Equation~\eqref{eq:diff-CK} must be complemented with appropriate
boundary conditions. If the movement takes place over the whole real
line, we have that $\lim_{x\to\pm\infty}\Pc=0$. If the movement
takes place over a finite interval $[-L,L]$ with periodic boundary
conditions, we have
\begin{align}
  P(-L,c,t)&=P(L,c,t),  & J(-L,c,t)&=J(L,c,t).
 \label{eq:periodic-bc}
\end{align}
The case with periodic boundary conditions is especially relevant for
the feedback flashing ratchet, in which the potential $V(x,c)$ is
spatially periodic. See~\ref{app:periodic-bc}
for a brief discussion on the relation between the problems over the
whole real line and the problem with periodic boundary conditions.

The evolution equation for $\Pc$ can be
compactly written in operator form
\begin{align}
  \partial_{t}P=&\hat{\calL}^{(c)}P+
  \sum_{k}\delta(t-t_{k}) \hat{W}P,
  \label{eq:joint-evolution}
\end{align}
where $\hat{\calL}^{(c)}$ is the Fokker-Planck operator for the
potential $V(x,c)$,
\begin{align}
  \label{eq:LC-def}
  \hat{\calL}^{(c)}P\equiv -\partial_{x} \J,
\end{align}
and $\hat{W}$ is the master equation operator for the control updates
\begin{equation}\label{eq:W-def}
  \hat{W}P\equiv \sum_{k}\Big[
  \hat{\Theta}_{c}\sum_{c'\ne c}P(x,c',t) -(I-\hat{\Theta}_{c})\Pc\Big],
\end{equation}
where we have defined linear operators $\hat{\Theta}_{c}$,
\begin{subequations}\label{eq:theta-C-op}
\begin{align}
  &\hat{\Theta}_{c}f(x)\equiv\Theta(c|x)\, f(x), \quad \forall f(x),
  \\
  &\sum_{c}\hat{\Theta}_{c}=I.
\end{align}
\end{subequations}
In terms of them, Eq.~\eqref{eq:P-tk-+} can be written as
\begin{equation}
  \label{eq:P-tk-+-v2}
  P(x,c,t_{k}^{+})= \hat{\Theta}_{c} P(x,t_{k}).
\end{equation}
For error-free measurements, it is clear that the operators
$\hat{\Theta}_{c} $ are projectors, verifying
$\hat{\Theta}_{c}^{2}=\hat{\Theta}_{c}$. Their action on a function
$f(x)$ is to restrict it to the interval $\calX_{c}$, cutting its
remainder.

\subsection{Formal solution of the differential Chapman-Kolmogorov equation~\eqref{eq:diff-CK}}\label{sec:formal-sol}

Our general framework allows us to write a formal solution of the DCKE~\eqref{eq:diff-CK}, Eq.~\eqref{eq:joint-evolution} in operator form. In the intervals $I_{k}$, the joint distribution $\Pc$ evolves following the Fokker-Planck equation with potential $V(x,c)$: denoting the evolution operator for such Fokker-Planck equation by $\hat{U}^{(c)}(t)\equiv
\exp(t\hat{\calL}^{(c)})$, 
\begin{equation}
  \label{eq:TC-t-FP}
 \Pc=\hat{U}^{(c)}(t-t_k)P(x,c,t_k^+), \quad t \in I_{k}.
\end{equation}
More specifically, in terms of the Green function $K(x,\Delta t| x',c)$ of the Fokker-Planck operator $\hat{\calL}^{(c)}$:
\begin{align}\label{eq:evol-fp}
    \Pc=\int dx' K(x,t-t_k^+| x',c)\, P(x',c,t_k^+), \quad t\in I_{k}.
\end{align}
In particular, at the time $t_{k}^-$ just before the measurement,
\begin{align}\label{eq:evol-fp-t-k+1}
    P(x,c,t_{k}^-)=\int dx' K(x,\Delta t_m| x',c) \,P(x',c,t_{k-1}^+).
\end{align}
At the measurement times, we resort to Eq. \eqref{eq:P-tk-+},
\begin{align}\label{eq:evol-measure}
    P(x,c,t_k^+)=\Theta(c|x) \sum_{c'}\int dx'\; K(x,\Delta t_m| x',c') P(x',c',t_{k-1}^+).
\end{align}

The Green function can be written as an expansion in the eigenfunctions of the Fokker-Planck operator $\hat{\calL}^{(c)}$. We define the eigenvalues and eigenfunctions of $\hat{\calL}^{(c)}$:
\begin{equation}
    \label{eq:FP-eigenfunctions}
    \hat{\calL}^{(c)} P_{\sts}^{(c)}=0, \qquad \hat{\calL}^{(c)} \phi_n^{(c)}=-\lambda_n^{(c)} \phi_n^{(c)}, \quad \lambda_n^{(c)}>0.
\end{equation}
Therefore, $\PsXC$ is the stationary distribution of the particle
moving under the action of the potential $\tilde{V}(x,c)$. For the case of the feedback flashing ratchet, this distribution is computed in~\ref{app:stationary-sol}. We recall that the Fokker-Planck operator $\hat{\calL}^{(c)}$ is Hermitian when the scalar product includes $1/P_{\sts}^{(c)}$ as a weight function~\cite{gardiner_stochastic_2009,van_kampen_stochastic_1992}, which makes it possible to write the Green function as
\begin{equation} \label{eq:green-function}
    K(x,t|x',c)=\PsXC+\frac{1}{\PsXprimeC}\sum_{n=1}  \phi^{(c)}_n(x')\phi^{(c)}_n(x) e^{-\lambda^{(c)}_n t}.
\end{equation}
This completes the formal solution of the problem: concatenating Eqs.~\eqref{eq:evol-fp} and \eqref{eq:evol-measure}, with the Green function being given by Eq.~\eqref{eq:green-function}, $\Pc$ for any $t$ is written in terms of the initial condition $P(x,c,t=0)$.

\section{$H$-theorem and long-time regime}\label{sec:H-theorem}

In this section, we derive an H-theorem that guarantees the convergence of the system to a unique long-time state. We start by considering two solutions $\Pc$ and $\Pper$ of the DCKE~\eqref{eq:diff-CK} and defining the  functional
\begin{equation}
    H(t)=\sum_c\int_{-\infty}^{+\infty} dx \Pc \ln \frac{\Pc}{\Pper}. 
\end{equation}
Note that this functional is the Kullback-Leibler divergence or the relative entropy. In the following, we prove that $H$ is a Lyapunov functional for the DCKE~\eqref{eq:diff-CK}, which implies that all its solutions tend to a unique asymptotic behaviour for long enough times, i.e.
\begin{equation}
    \lim_{t\to\infty} \frac{\Pc}{\Pper}=1.
\end{equation}
For the sake of concreteness, the proof is presented for periodic boundary conditions, as given by Eq.~\eqref{eq:periodic-bc}.
  
\subsection{$H$ is a decreasing function of time}\label{sec:H-decreasing}

Let us prove the $H$-theorem in two steps. First, we explore the behaviour of $H$ in the time interval $I_k$ between control updates. Second, we investigate how $H$ changes at the measurement times, i.e. $H(t_k^+)-H(t_k^-)$.

\subsubsection{During $I_k$.-}\label{sec:H-dec-Ik}

Taking the time derivative of $H$, we have
\begin{align}
    \dH
    = & \sum_c \int dx \, \partial_t \Pc \ln \frac{\Pc}{\Pper} 
     -\sum_c \int dx\, \frac{\Pc}{\Pper} \partial_t \Pper \nonumber \\
    = & -\sum_c \int dx \, \partial_x \J \ln \frac{\Pc}{\Pper} +\sum_c \int dx\, \frac{\Pc}{\Pper} \partial_x \Jper
\end{align}
Integrating by parts and bringing to bear the periodic boundary conditions \eqref{eq:periodic-bc}, we get
\begin{align}
  \dH
    = & \sum_c \int dx \, \Pc  \left[ \frac{\J}{\Pc}  - 
    \frac{\Jper }{\Pper} \right] \partial_x \ln \frac{\Pc}{\Pper} \label{eq:H-th-1} 
\end{align}
Equation~\eqref{eq:J-def} is equivalent to
\begin{equation}\label{eq:part-x-ln-P}
    \partial_x\ln \Pc=-\beta \left(\gamma \frac{\J}{\Pc}+c V'(x)+F_{\ext} \right),
\end{equation}
and an analogous equation for $\partial_x\ln \Pper$ and $\Jper$. Therefore, we have
\begin{equation}\label{eq:difference-of-Js}
    \frac{\J}{\Pc} -\frac{\Jper }{\Pper}=-(\beta \gamma)^{-1} \partial_x \ln \frac{\Pc}{\Pper} .
\end{equation}
Inserting this equation into Eq.~\eqref{eq:H-th-1}, we get 
\begin{align}
     \dH &= -(\beta \gamma)^{-1} \sum_c \int dx \, \Pc \left(\partial_x \ln \frac{\Pc}{\Pper}\right)^2. 
     \label{eq:H-th-2} 
\end{align}
By defining 
\begin{equation}\label{eq:sigma-C}
     \sigma_c(t)\equiv (\beta \gamma)^{-1} \int dx \, \Pc \left(\partial_x  \ln \frac{\Pc}{\Pper}\right)^2 \geq 0,
\end{equation}
we rewrite Eq.~\eqref{eq:H-th-2} as
\begin{equation}
    \dH=- \sum_c \sigma_c(t)\le 0.
\end{equation}

Since $\sigma_c(t)\ge 0$, $\forall c$,  it is clear that $\dot{H}=0$ implies that $\sigma_c(t)$ must vanish, $\forall c$, and this in turn implies that $\Pc=k(c,t) \Pper$. Integrating over $x$, $P(c,t)=k(c,t) \tilde{P}(c,t)$,
which is valid between control updates, i.e. for $t\in I_k$. But neither $P(c,t)$ nor $\tilde{P}(c,t)$ depends on time between control updates, which entails that $k(c,t)$ only depends on $c$ throughout $I_k$. Therefore, we conclude that 
\begin{equation}\label{eq:k(c)}
     \dot{H}=0 \text{ for } t\in I_k \iff \Pc=k(c) \Pper.
\end{equation}

\subsubsection{At the control updates.-}\label{sec:H-change-tk}
 
{Let us compute the change of the $H$ functional at the control updates, i.e. 
\begin{equation}
    \delta_k\equiv H(t_k^+)-H(t_k^-).
\end{equation}
On the one hand,
\begin{align}
     H(t_k^+)&= \sum_c \int dx\, \Pca \ln \frac{\Pca}{\Ppera}
     = \int dx\, \Pb \ln \frac{\Pb}{\Pperxb}  
\end{align}
where we have brought to bear Eq. \eqref{eq:P-tk-+}.
On the other hand, 
\begin{align}
     H(t_k^-)&= \sum_c \int dx \Pcb \ln \frac{\Pcb}{\Pperb}  .
\end{align}
Therefore, we have that
\begin{align}
    \delta_k&=\sum_c \int dx\,  \Pcb \ln \frac{\Pb}{\Pperxb} -\sum_c \int dx \,  \Pcb \ln \frac{\Pcb}{\Pperb} \nonumber \\
     &= -\sum_c \int dx \,\Pcb \ln \frac{\Pcb\Pperxb}{\Pb\Pperb} \nonumber \\
     &= -  \int dx \, \Pb \sum_c P(c|x,t_k^-) \ln \frac{P(c|x,t_k^-)}{\tilde{P} (c|x,t_k^-)} \le 0.
\end{align}
To ensure that $\delta_k \leq 0$, it is convenient to introduce the rewriting 
\begin{align}
    \sum_c P(c|x,t_k^-) \ln &\frac{P(c|x,t_k^-)}{\tilde{P} (c|x,t_k^-)}= \sum_c \tilde{P} (c|x,t_k^-) \left[ \frac{P(c|x,t_k^-)}{\tilde{P} (c|x,t_k^-)} \ln \frac{P(c|x,t_k^-)}{\tilde{P} (c|x,t_k^-)} + 1-\frac{P(c|x,t_k^-)}{\tilde{P} (c|x,t_k^-)}\right].
\end{align}
Now we take into account that $x \ln x +1-x \geq 0$, with the equality holding if and only if $x=1$. Then, by defining $x={P(c|x,t_k^-)}/{\tilde{P} (c|x,t_k^-)}$, we conclude that $\delta_k\le 0$, being equal to zero if and only $P(c|x,t_k^-)=\tilde{P} (c|x,t_k^-)$ or, equivalently, 
\begin{equation}\label{eq:PC-cond-x-tk}
    \delta_k=0 \iff P(c|x,t_k^-)=\tilde{P}(c|x,t_k^-), \; \forall (x,c).
\end{equation}
}

The above discussion entails that $H$ is discontinuous at the control updates, for the wide class of feedback controlled systems we are considering. Yet, it must be noted that the discontinuity of $H$ persists for open-loop controllers, in which the control does not use information about the system state, since in general $P(c|x,t_k^-)\ne \tilde{P}(c|x,t_k^-)$.

\subsection{Long-time limit and tendency towards a universal behaviour}

In the previous sections, we have shown that $H$ is non-increasing both during the time intervals $I_k$ and the time instants $t_k$ at which the control is updated. Here we prove that, since $H$ is bounded from below, this means that (i) $\dot{H}\to 0$ for $t\in I_k$, $t_k^+<t<t_{k+1}^-$, and (ii) $\delta_k=H(t_k^+)-H(t_k^-)\to 0$, for large enough $k$. 

The total decrease of the $H$-functional would be
\begin{equation}
    \Delta H =\sum_{k=1}^{\infty} \underbrace{\delta_k}_{\le 0} + \sum_{k=0}^{\infty} \underbrace{\int_{t_k^+}^{t_{k+1}^-} dt \, \dH}_{\equiv\varepsilon_k\le 0}\le 0 .
\end{equation}
Since $H$ is bounded from below, the two series on the rhs must be convergent, which implies that 
$$\lim_{k\to\infty}\delta_k=0, \quad \lim_{k\to\infty}\varepsilon_k=0.$$ 
The second condition, $\varepsilon_k\to 0$ for $k\to\infty$ is equivalent to stating that $\dot{H}\to 0$ over the time interval $I_k$, for large enough $k$. Let us consider Eq.~\eqref{eq:k(c)} at $t=t_k^-$, the condition $\dot{H}=0$ entails that
\begin{align}
    \Pcb=k(c) \Pperb.
\end{align}
Once more, we resort to the Bayes rule to write
\begin{align}
    \cancel{P(c|x,t_k^-)}\Pb=k(c) \cancel{\tilde{P}(c|x,t_k^-)}\Pperxb ;
\end{align}
the cancelled terms on both sides follow from the condition $\delta_k=0$, as expressed by Eq.~\eqref{eq:PC-cond-x-tk}. Integrating over all $x$, we obtain
\begin{equation}
    k(c)=1 \implies \Pc=\Pper, \; \forall t\in I_k,
\end{equation}
bringing to bear again Eq.~\eqref{eq:k(c)}.\footnote{The equality of the joint probabilities $\Pc$ and $\Pper$ at $t=t_k^-$ entails that the conditional probabilities in Eq.~\eqref{eq:PC-cond-x-tk} are also equal, consistently with Eq.~\eqref{eq:PC-cond-x-tk}.}

To conclude, we have proven that
\begin{align}
    \lim_{t\to\infty} \frac{\Pc}{\Pper}=1 \quad \text{or}  \quad 
  \Pc \sim \Pper, \; t\to +\infty.
\end{align}
That is, all the solutions of the differential Chapman-Kolmogorov equation~\eqref{eq:diff-CK} tend towards a universal behaviour, independent of the initial conditions, for long enough times. It is this universal long-time regime that we are especially interested in.

\section{Coarse-graining: Markovian equation for the position}\label{sec:coarse-grain-Markov-x}

Just after the control updates, the conditional probability of $c$
for given $x$ is determined by the function $\Theta(c|x)$, which
characterises the action of the control---as expressed by
Eq.~\eqref{eq:P-cond-C-on-x}.  {First, in Sec.~\ref{sec:markov-position}, we take advantage of this fact to write down a closed equation for the marginal distribution $P(x,t_{k+1})$ in terms of
$P(x,t_{k})$, i.e. to derive a Markov chain in the coarse-grained times $t_k=k\Delta t_m$. Interestingly, it is not necessary that the measurement is error-free: the Markovian description for the position in this coarse-grained timescale is also valid for measurements involving a given uncertainty $\Delta x$, as those in Eq.~\eqref{eq:theta-c-imperfect}. Second, in Sec.~\ref{sec:nonmarkov-control}, we show that the marginal process for the control is not Markovian, but its distribution can be inferred from that of the position. Third, in Sec.~\ref{sec:mutual-info}, we discuss the mutual information between the control and the position. Finally, in Sec.~\ref{sec:relevant-limits}, we analyse two relevant limit situations, those of very frequent and very infrequent control updates, formally $\Delta t_m\to 0^+$ and $\Delta t_m\to\infty$. Therein, explicit analytical expressions for the long-time behaviour of the probability distributions are derived.}

\subsection{Markov chain for the position}\label{sec:markov-position}

We introduce the notation
\begin{equation}\label{eq:rho-k-def}
  P^{(k)}(x)\equiv P(x,t_{k})
\end{equation}
for the position distribution at the measurement times---recall that the marginal distribution $P(x,t)$ is continuous thereat. Marginalising Eq.~\eqref{eq:evol-fp-t-k+1}, we get
\begin{align}
P^{(k)}(x)&
= \sum_c\int dx'\, K(x,\Delta t_{m}|x',c) P(x',c,t_{k-1}^+) \nonumber \\ 
&=\sum_c\int dx'\,K(x,\Delta t_{m}|x',c) \Theta(c|x') P^{(k-1)}(x'),
 \label{eq:TC-t(k+1)-FP}
\end{align}
where we have also made use of \eqref{eq:P-tk-+}. Formally, we can write
\begin{equation}
\label{eq:TC-t(k+1)-FP-op}
   P^{(k)}(x)= \Big[\sum_c \hat{U}^{(c)}(\Delta t_{m})\hat{\Theta}_{c}\Big]P^{(k-1)}(x)
\end{equation}
or, more specifically,
\begin{subequations} \label{eq:TC-t(k+1)-FP-explicit-v2}
\begin{align}
  P^{(k)}(x)&=\int dx'\, T(x,\Delta t_{m}|x')P^{(k-1)}(x'), \\
  T(x,\Delta t_{m}|x')&\equiv \sum_{c} K(x,\Delta t_{m}|x',c)\Theta(c|x').
\end{align}
\end{subequations}
Of course, Eq.~\eqref{eq:TC-t(k+1)-FP-explicit-v2} conserves the total probability: since the
Green function 
conserves the spatial integral of the function to its right, $\int dx\,P^{(k)}(x)=\int dx\,P^{(k-1)}(x)$.

Equation~\eqref{eq:TC-t(k+1)-FP-explicit-v2} defines  a Markov chain for the distribution of the position at the discrete times of the control updates. Our
coarse-graining of time has made $X$ Markovian, whereas the joint process
$(X,C)$ is Markovian over the ``finer'' timescale, which includes both the control updates and the time evolution between them. Over this fine timescale, $X$ is not Markovian; it is only over a ``rougher'' timescale that $X$ alone becomes a Markov process---in complete analogy with the situation in other
physical contexts, such as Brownian motion~\cite{gardiner_stochastic_2009,van_kampen_stochastic_1992}.

Our (continuous) stochastic matrix $T(x,\Delta t_{m}|x')$ has
a left eigenvector with all its components equal to unity: if
$\phi(x)\equiv 1$, $\forall x$, 
\begin{align}
  \int dx\,  \phi(x) T(x,\Delta t_{m}|x')=\sum_{c}\Theta(c|x')=1=\phi(x'),
\end{align}
since $\int dx\, K(x,t|x',c)=1$, $\forall (t,x',c)$. As a consequence,
there should be a right eigenvector corresponding to the unit
eigenvalue, i.e.~a time-independent solution $P_{\sts} (x)$,
\begin{equation}
  \int dx' \, T(x,\Delta t_{m}|x') P_{\sts} (x')= P_{\sts} (x).
  \label{eq:Ps-x-int-eq}
\end{equation}
On physical grounds, we expect $P_{\sts} (x)$ to be a meaningful
probability distribution, with $P_{\sts} (x)\ge 0$ $\forall x$, when
working in the finite interval $x\in[-L,L]$. On the whole real
line, there would not be in general a normalisable stationary
distribution.\footnote{The non-negativity of
$P_{\sts}(x)$ could be also argued on the basis of the Perron-Frobenius theorem, but rigorously this theorem holds for a finite set of discrete states, not for an infinite set with a continuous ``index'' $x$ as we have here.}

On the one hand, the $H$-theorem for the joint process $(X,C)$ over the fine timescale shows that the system reaches a long-time ``normal" state~\cite{brey_normal_1993}, independent of initial conditions. Physically, one expects this solution to be time-periodic, with period $\Delta t_{m}$. Over the rougher timescale for which $X$ becomes
Markovian, we have only the instants just after the control is updated,
which are precisely separated by $\Delta t_{m}$. In this description,
the long-time behaviour of $P^{(k)}(x)$ should thus correspond to a
stationary, time-independent, state. This reinforces our expectation
of having a meaningful solution of Eq.~\eqref{eq:Ps-x-int-eq}.

On the other hand, as shown in Sec.~\ref{sec:formal-sol}, it is necessary to sequentially concatenate the action of the Fokker-Planck and measurement operators to obtain $\Pc$ from a given initial condition. However, if we are just interested in the long-time state, and not in the transitory relaxation thereto, the stationary marginal distribution $P_{\sts}(x)$ makes it possible to derive the long-time state for the joint probability $\Pc$. By combining Eqs.~\eqref{eq:evol-fp} and \eqref{eq:P-tk-+}, particularised for the steady marginal distribution $P_{\sts}(x)$, we get
\begin{equation}\label{eq:P-period}
    P_{\period}(x,c,t)=\int dx'\; K(x,t-t_k^+| x',c) \Theta(c|x')P_{\sts}(x'), \quad t\in I_k.
\end{equation}
Note that, by construction, $P_{\period}(x,c,t)$ is indeed a periodic solution, with period $\Delta t_m$, of the DCKE~\eqref{eq:diff-CK}. Our $H$-theorem ensures that the joint process $(X,C)$ monotonically approaches this periodic solution for long enough times. This is an interesting point, since we are able to compute the long-time regime in the finer time scale, which is non-stationary, from the steady state that emerges in the coarse-grained  scale.

\subsection{Non-Markovianity of the control}\label{sec:nonmarkov-control}

Now we consider the time evolution of the marginal probability for the
control $P(c,t)$ over this coarse-grained timescale, and show that it
is not Markovian. The marginal probability of the control is
\begin{equation}\label{eq:p-k-def}
  P^{(k)}(c)\equiv P(c,t^+_{k}).
\end{equation}
Over each time interval $t\in I_{k}$, the
probability of the control does not evolve and is given by $P^{(k)}(c)$. Notice that, unlike in the case of the position, this marginal probability is not continuous at the control updates, in general $P(c,t_k^-)\neq P(c,t_k^+)$.  By recalling Eqs.~\eqref{eq:P-tk-+} and~\eqref{eq:rho-k-def}, we have
\begin{equation}
  \label{eq:P-k-C}
 P^{(k)}(c)=\int dx \,P^{(k)}(x)\Theta(c|x).
\end{equation}

The evolution of $P^{(k)}(c)$ is obtained from that of $P^{(k)}(x)$. Taking into account Eq.~\eqref{eq:TC-t(k+1)-FP-explicit-v2}, we get
\begin{align}
  P^{(k+1)}(c)= \sum_{c'}\int \!dx \! \int \!dx' 
&\Theta(c|x) \Theta(c'|x')  K(x,\Delta t_{m}|x',c')P^{(k)}(x'),
  \label{eq:P-k+1-C}
\end{align}
which cannot be cast into a closed equation for the evolution of
$P^{(k)}(c)$, due to the dependence on $x'$ of
$K(x,\Delta t_{m}|x',c')$. It is only in the limit as
$\Delta t_{m}\to\infty$ that $K(x,\Delta t_{m}|x',c')$ becomes
independent of $x'$ and $P^{(k)}(c)$ obeys a closed evolution
equation---see Sec~\ref{sec:delta-tm-infty}. Yet, the tendency of $P^{(k)}(x)$ to a steady state implies that also
$P^{(k)}(c)$ tends to a steady distribution
$P_{\sts}(c)$. Equation~\eqref{eq:P-k-C} tells us that
\begin{equation}
  \label{eq:P-k-C-sts}
  P_{\sts}(c)=\int dx \,P_{\sts}(x)\,\Theta(c|x),
\end{equation}
the stationary probability of each value of $c$ is given, logically,
by the steady probability of finding the position of the particle in
the corresponding interval.

The physical picture stemming from the above discussion is appealing.
Over the coarse-grained timescale, $X$ becomes a Markov process but
$C$ is not: $C$ does not have enough information. If you know the
position just after the control update, you know the conditional
probability of the control---but not the other way round.

\subsection{Mutual information}\label{sec:mutual-info}

In our model, the feedback controller {\color{black} measures the position of the particle} at the regularly spaced times $t_k$. {\color{black}The measurement thus correlates the position $X$ and the controller variable $C$, and this correlation is quantitatively measured by the mutual information between these variables just after these times, i.e. at $t=t_k^+$. The mutual information} is defined as~\cite{cover_elements_2006} 
\begin{equation}
   I_k(X,C)\equiv \sum_c \int dx \, P(x,c,t_k^+) \ln {\frac{P(x,c,t_k^+)}{P(x,t_k^+)P(c,t_k^+)}}=\sum_c \int dx \, P(x,c,t_k^+) \ln {\frac{P(c|x,t_k^+)}{P(c,t_k^+)}} 
   \label{eq:Is-def}
\end{equation}
where we have used Eqs. \eqref{eq:P-tk-+} and \eqref{eq:P-tk-+-marginal-continuity}. We recall that $I_k(X,C)$ is non-negative, and it equals zero if and only if $X$ and $C$ are independent at $t=t_k^+$. Now, we can make use of Eq.~\eqref{eq:PC-cond-x-tk} and the tendency of $P(x,t_k^+)$ to a steady state solution $P_{\sts}(x)$ for large $k$ to write an expression for the  mutual information in the long-time regime:
\begin{equation}
    I_k(X,C) \stackrel{k\gg 1}{\longrightarrow} I_{\sts}(X,C), \qquad  I_{\sts}(X,C)\equiv \sum_c \int dx \, \Theta(c|x) P_{\sts}(x) \ln {\frac{ \Theta(c|x)}{P_{\sts}(c)}}.
\end{equation}
Moreover, bringing to bear Eq.~\eqref{eq:P-k-C-sts}, we have that
\begin{align}
    I_{\sts}(X,C) &=-\sum_c P_{\sts}(c)\ln{P_{\sts}(c)} +\int  dx \, P_{\sts}(x) \sum_c \Theta(c|x) \ln{\Theta(c|x)} .
\label{eq:mutual-information-all-tm}
\end{align}

{
Note that the mutual information can thus be calculated from the marginal probability $P_{\sts}(x)$ over the coarse-grained timescale and the protocol for updating the control value $\Theta(c|x)$---recall that $P_{\sts}(c)$ is given in terms of $P_{\sts}(x)$ by Eq.~\eqref{eq:P-k-C-sts}. The second term on the rhs is always non-positive and, therefore, the maximum value of $I_{\sts}(X,C)$ is the (dimensionless) stationary entropy of the controller. This maximum value is attained for the perfect closed-loop case, for which the second term of Eq.~\eqref{eq:mutual-information-all-tm} vanishes. The mininum value is attained for open-loop control: in that case, $\Theta(c|\cancel{x})=\Theta(c)$ does not depend on $x$ and Eq.~\eqref{eq:P-k-C-sts} tells us that $P_{\sts}(c)=\Theta(c)$, which leads to $I_\sts(X,C) =0$.
}

In Sec. \ref{analytical-results}, we particularise the above expressions to the case of the feedback flashing ratchet. Also therein, we discuss the relation between the {\color{black} mutual} information and the induced current.

\subsection{Relevant limits}\label{sec:relevant-limits}

\subsubsection{Limit $\Delta t_{m}\to\infty$.-}\label{sec:delta-tm-infty}

Let us analyse the case in which the frequency of control updates is
very low, $\Delta t_{m}\to\infty$. Physically, this limit appears
when the time between control updates is much longer than all the rest
of the characteristic times of the system.

In this situation, the Green function verifies
\begin{equation}
  K(x,\Delta t_{m}\to\infty|x',c)\to \PsXC.
\end{equation}
Therefore, we have the following
equation for $P_{\sts}(x)$,
\begin{align}
  P_{\sts} (x)&=\sum_{c}\left(\int dx' \Theta(c|x')
                   P_{\sts} (x')\right) \PsXC 
  =\sum_{c}P_{\sts}(c) \PsXC.
  \label{eq:Ps-deltatm-inf}
\end{align}
where we have made use of Eq.~\eqref{eq:P-k-C-sts}. The stationary solution $P_{\sts}(x)$ is a weighted average of the
steady solutions $\PsXC$ for the problems with the different potentials
$\tilde{V}(x,c)$.

To solve Eq.~\eqref{eq:Ps-deltatm-inf} for $P_{\sts}(x)$, it is, therefore, necessary to derive an equation for $P_{\sts}(c)$. It is handy to introduce a matrix $B$ with elements
\begin{equation}
  \label{eq:matrixB-def}
  B_{cc'}\!\equiv\int dx \, \Theta(c|x)P_{\sts}^{(c')}(x).
\end{equation}
In the limit $\Delta t_{m}\to\infty$ we are considering, Eq.~\eqref{eq:P-k+1-C} can be rewritten as
\begin{equation}
  P^{(k+1)}(c)=\sum_{c'}B_{cc'}P^{(k)}(c'),
  \label{eq:Markov-chain-C}
\end{equation}
which is the evolution equation of a Markov chain. As already commented in the previous section, it is only in this limit that the marginal process $C$ becomes Markovian.  The matrix $B$ with elements $B_{cc'}$ is its corresponding \textit{stochastic
  matrix}, such that
\begin{equation}
  B_{cc'}>0, \quad \sum_{c}B_{cc'}=1.
\end{equation}
The matrix $B$ has thus the unit eigenvalue and, by virtue of the Perron-Frobenius theorem, the
corresponding right eigenvector, which verifies
\begin{equation}\label{eq:Psc-tm-infty}
    P_{\sts}(c)=\sum_{c'}B_{cc'}P_{\sts}(c'),
\end{equation}
is unique and  all its components $P_{\sts}(c)$ are
positive.\footnote{This equation can also be obtained from~\eqref{eq:Ps-deltatm-inf}, multiplying it by $\Theta(c|x)$ and integrating over $x$.} Normalising this eigenvector in the usual sense, we have
\begin{equation}
  P_{\sts}(c)>0, \quad \sum_{c}P_{\sts}(c)=1,
\end{equation}
which ensures that $P_{\sts}(x)$ in
Eq.~\eqref{eq:Ps-deltatm-inf} is a meaningful probability
distribution, moreover correctly normalised.

{Therefore, in the limit $\Delta t_m\to\infty$, by solving Eq.~\eqref{eq:Psc-tm-infty} we find the stationary solution for the marginal distribution of the control $P_{\sts}(c)$. Inserting this result into Eq.~\eqref{eq:Ps-deltatm-inf}, we get the stationary solution of the marginal distribution of the position $P_{\sts}(x)$ over the coarse-grained timescale. In turn, from our knowing of $P_{\sts}(x)$, we get the periodic long-time state $P_{\period}(x,c,t)$ of the joint probability distribution, using Eq.~\eqref{eq:P-period}.}

\subsubsection{Limit $\Delta t_{m}\to 0^+$.-}\label{sec:delta-tm-zero}

Another relevant physical situation is that of a high frequency of
control updates, i.e. $\Delta t_{m}\to 0^+$. Physically, this limit appears
when the time between control updates is much shorter than all the rest
of the characteristic times of the system.

In this limit, it is useful to introduce the short-time expansion
\begin{equation}
  \label{eq:Uc-short-time}
  \hat{U}^{(c)}(\Delta t_{m})=I+\Delta t_{m}\hat{\calL}^{(c)}.
\end{equation}
Then, Eq.~\eqref{eq:TC-t(k+1)-FP-op} is approximated as
\begin{align}
  P(x,t_k)&=
  \sum_{c}\left[I+\Delta t_{m}\hat{\calL^{(c)}}\right]\hat{\Theta}_{c}P(x,t_{k-1})
                =P(x,t_{k-1})+\Delta t_{m}
                \sum_{c}\hat{\calL^{(c)}}\hat{\Theta}_{c}P(x,t_{k-1}),
                \label{eq:TC-t(k+1)-FP-short-time}
\end{align}
which can be cast as a first-order differential equation
in time, recalling that $t_{k}=k\Delta t_{m}$,
\begin{equation}
  \label{eq:TC-t(k+1)-FP-short-time-diff-op}
  \partial_{t}P(x,t)=
  \sum_{c}\hat{\calL^{(c)}}\hat{\Theta}_{c}P(x,t),
\end{equation}
or
\begin{align}
  \gamma \,\partial_{t}P(x,t)&=
  \sum_{c} \partial_{x}\Big[\tilde{V}'(x,c)\,
                      \Theta(c|x)P(x,t) 
   +k_{B}T\,\partial_{x}(\Theta(c|x)P(x,t)) \Big],
\end{align}
{ where we use the prime to indicate the spatial derivative. By defining the effective potential and force
\begin{subequations}\label{eq:effective-pot-short-times}
\begin{align}
    V_{\eff}(x)&\equiv \sum_{c}\int_{0}^{x} dx' \,{V}'(x',c)\Theta(c|x'), \\
    -V'_{\eff}(x)&=-\sum_{c}{V}'(x,c)\Theta(c|x),
\end{align} 
\end{subequations}
we have that
\begin{equation}
    \gamma\,\partial_t P(x,t)= \partial_{x}\Big[\tilde{V}'_{\eff} (x)P(x,t)+k_{B}T\,\partial_{x}P(x,t)\Big],
  \label{eq:TC-t(k+1)-FP-short-time-diff-v2}
\end{equation}
where $\tilde{V}'_{\eff}=V'_{\eff}+F_{\ext}$.
}

Equation~\eqref{eq:TC-t(k+1)-FP-short-time-diff-v2} is the
Fokker-Planck equation for an overdamped particle moving under the
action of a conservative force field $-V'_{\eff}(x)$ plus
the non-conservative force $-F_{\ext}$. In the long-time limit, the
system reaches the corresponding steady solution {$P_{\sts}(x)$, which verifies
\begin{equation}
    \tilde{V}'_{\eff}(x)P_{\sts}(x)+\beta^{-1}P'_{\sts}(x)=\text{const}.
\end{equation}
See~\ref{app:stationary-sol}} for details, in particular Eq.~\eqref{eq:Ps-sol-v2}. For error-free measurements, the effective potential $V_{\eff}(x)$
can be seen as the continuous concatenation of the different
potentials $V(x,c)$, each one restricted to the region $\calX_{c}$ for
which $\Theta(c|x)=1$. This may be used, as in the feedback flashing
ratchet considered below, to generate a directed motion in a certain
direction.

\section{Feedback flashing ratchet}\label{sec:flashing-ratchet}

\subsection{Description of the system}

In order to illustrate the formalism developed in the previous sections, now we apply it to a specific system: the feedback flashing ratchet sketched as a motivation for our work in Sec.~\ref{sec:motivation-feedback-ratchet}. To be concrete, we consider the particle to move under the action of a V-shaped potential in the absence of external force:
\begin{equation}\label{eq_V-shape}
    V(x)=F \abs{x}, \quad  x \in [-L, L], \qquad F_{\ext}=0.
\end{equation}
The feedback controller switches on or off the potential $V(x)$ depending on the particle position $x(t_k^-)$ just before the measurement time, as described in Eq. \eqref{eq:theta-c-imperfect}. Therefore, the fluctuating potential $V(x,c)$ introduced in Sec.~\ref{sec:general-frame} is
\begin{equation}
    V(x,c)=c\,F\abs{x},
\end{equation}
and the corresponding conservative  force  acting on the particle is 
\begin{equation}
    -\partial_x V(x,c)=-c\, F \sgn(x)=c \, F \sgn(-x).
\end{equation}
Therefore, the sets $\calX_1$ and $\calX_0$ defined by respectively having a positive and negative value of the force are given by $[-L,0)$ and $(0,L]$---see the right panel of Fig.~\ref{fig:V-shaped-potential}.

The most general formulation of the mathematical problem for the specific system we are considering  can be written as follows. We have to solve the DCKE~\eqref{eq:diff-CK} for the joint probability $P(x,c,t)$, particularised for the feedback flashing ratchet with the V-shaped potential:
\begin{align} 
    \partial_t \Pc=& 
    \gamma^{-1}  c \, F \,\partial_x \left[\,\sgn(x)  \Pc\right]+(\gamma \beta)^{-1} \partial_x^2 \Pc\nonumber \\
    &+ \sum_k \delta (t-t_k^-) \Big[\Theta(c|x)P(x,1-c,t')-\Theta(1-c|x)\Pc \Big],
\label{eq:flashing-problem}
   \end{align}
with the feedback control protocol that takes into account an uncertainty in the particle position $\Delta x$:
   \begin{align} \label{eq:flashing-protocol}
    \Theta(c|x)=\begin{cases}
  \text{length\;} \{ [-L,0] \cap(x-\Delta x, x+ \Delta x)\}/(2 \Delta x) , & \text{if }c=1, \\
    \text{length\;} \{ [0,L] \cap (x-\Delta x, x+ \Delta x)\}/(2 \Delta x), & \text{if } c=0.
    \end{cases}
\end{align}
The error-free version of this protocol, for $\Delta x=0$, switches on (off) the potential if the force at the instantaneous position $x$ points to the right (left). For $\Delta x\ne 0$, we have localised the position of the particle in the interval $(x-\Delta x,x+\Delta x)$, so the potential is switched on (off) with probability proportional to the length of this segment that lies inside the interval $\calX_1$ ($\calX_0$) in which the force points to the right (left). The explicit expressions for $\Theta(c|x)$ can be found in~\ref{app:f-f-protocol}.

The boundary conditions for this problem are periodic, as given by Eq.~\eqref{eq:periodic-bc}. For the initial conditions, we have chosen
\begin{align}\label{eq:initial-conditions}
  P(x,1,0)=0, \quad P(x,0,0)=(2L)^{-1},
\end{align}
which corresponds to having the potential switched off at $t=0$, with a flat distribution of the particles in the interval $[-L,L]$. Note that, in the notation introduced in Sec. \ref{sec:dynamic-eq}, $P(x,0,0)=P_{\sts}^{(0)}(x)$---i.e. it is the stationary distribution for the potential $V(x,0)=0$.

\subsection{Analytical results} \label{analytical-results}

For this system, we can analytically find the joint probability $\Pc$, solution of the problem over the fine timescale that contains both the control updates and the evolution between them. This is achieved by employing the formal solution derived in Sec.~\ref{sec:formal-sol}, i.e. by concatenating Eqs. \eqref{eq:evol-fp} and \eqref{eq:evol-measure}, since we can exactly calculate the Green function $K(x,c|x',t)$.\footnote{The expression for the Green function as an infinite series is not especially illuminating, it is thus relegated to~\ref{app:eigenvalues-and-eigenfunctions}.}

Over the coarse-grained timescale that arises for the times $t_k^+$ just after the control updates, the system reaches a steady state and we are able to derive additional results. For example, in the two limiting cases introduced in Sec. \ref{sec:coarse-grain-Markov-x}, $\Delta t_m \to \infty$ and $\Delta t_m \to 0^+$, we can exactly calculate the steady marginal distribution of the position $P_{\sts}(x)$ over this coarse-grained timescale.

\subsubsection{Limit $\Delta t_m \to\infty$.-}

In this limit, we have that  Green function $K(x,\Delta t_m \to \infty|x',c) \to \PsXC$, which is independent of $x'$---see Eq.~\eqref{eq:Psts_x|c} in \ref{app:stationary-sol}. In Sec. \ref{sec:delta-tm-infty}, we have taken advantage of this property to derive a closed system of equations for the marginal probability of the control at the steady state $P_{\sts}(c)$, given by Eq.~\eqref{eq:Ps-deltatm-inf}. For the case we are considering, we can derive explicit expressions for the coefficients $B_{cc'}$, defined in Eq.~\eqref{eq:matrixB-def}. By taking into account that
\begin{align}\label{eq:marginal-steady-state}
    \PsXzero=\frac{1}{2L}, \quad   
    \PsXone=\frac{\beta F}{2(1-e^{-\beta F L})} e^{-\beta F\abs{x}},
\end{align}
a straightforward calculation leads to 
\begin{equation}\label{eq:Bcc'-1/2}
    B_{cc'}=\frac{1}{2}, \quad \forall c,c'=0,1.
\end{equation}
This implies that
\begin{equation}
    P_{\sts}(c)=\frac{1}{2}, \quad \forall c=0,1,
    \label{pc_infty}
\end{equation}
and making use of Eq.~\eqref{eq:Ps-deltatm-inf}, we finally obtain the steady distribution of the position over the coarse-grained timescale: 
\begin{equation}\label{eq:sol-infty-free-protocol}
    P_{\sts}(x)=\frac{1}{2} \left[\PsXzero+\PsXone \right], \quad \Delta t_m\to\infty.
\end{equation}
This result is also independent of the uncertainty in the measurement $\Delta x$. In fact, the results in Eqs.~\eqref{eq:Bcc'-1/2}--\eqref{eq:sol-infty-free-protocol} apply to a broad class of flashing ratchets, see~\ref{app:computation-of-bcc}.

\subsubsection{Limit $\Delta t_m \to 0^+$.-}

\begin{figure}
    \centering
    \includegraphics[width=0.75\linewidth]{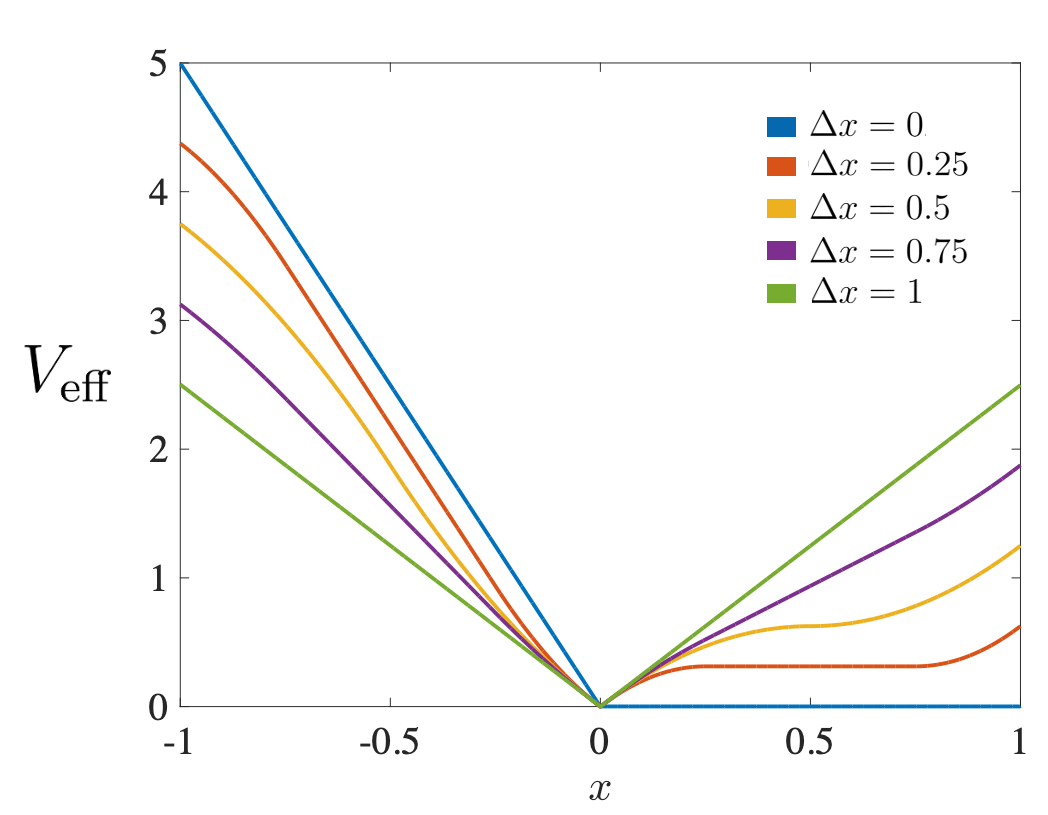}
    \caption{Effective potentials felt by the particle in the limit $\Delta t_m \to 0^+$, for different values of $\Delta x$. The colour code is the same as in Fig.~\ref{fig:protocols}. When $\Delta x=0$ (error-free protocol), the effective potential (blue line) corresponds to a constant positive force for $x<0$, and no force for $x>0$. As the uncertainty in the measurement $\Delta x$ increases, the protocol converges to the open-loop protocol and thus the effective potential tends to the V-shaped potential (green line), weighted by the probability $\Theta(1|\cancel{x})=1/2$ of switching it on.
}
    \label{fig:effective_potential}
\end{figure}

In this limit, we have shown in Sec.~\ref{sec:delta-tm-zero} that the particle effectively moves under the action of an effective potential. In the specific case we are considering, Eq.~\eqref{eq:effective-pot-short-times} particularises to
\begin{equation}\label{eq:Veff}
    V_{\eff}(x)=\int_0^x dx' \, \partial_{x'}V(x',1)\Theta(1|x')=F
    \int_0^x dx' \,  \sgn(x')
    \,\Theta(1|x'),
\end{equation}
since  the potential is switched off for $c=0$. The shape of this potential thus depends on uncertainty $\Delta x$ of the measurement, as illustrated in Fig.~\ref{fig:effective_potential}. The corresponding force is $-V'_{\eff}(x)= -V'(x) \Theta(1|x)=F\sgn(-x)\Theta(1|x)$, which reduces to $- V'_{\eff}(x)=F\eta(-x)$ for the error-free case $\Delta x=0$, where $\eta(x)$ is the Heaviside function. For any $\Delta x$, the stationary distribution for the position is 
\begin{equation}
\label{eq:Ps-zero-tm-protocol}
  P_{\sts}(x)=K \left[e^{-\beta V_{\eff}(x)}+ \frac{1-e^{-\beta [V_{\eff}(-L)-V_{\eff}(L)]}}{\int_{-L}^{L}dx'
  e^{\beta V_{\eff}(x') }  }\int_{-L}^{x}dx'
  e^{-\beta\left[V_{\eff}(x)-V_{\eff}(x')\right] }  \right] , 
\end{equation}
where $K$ is a normalisation constant---see \ref{app:stationary-sol}.

We recall that Eq.~\eqref{eq:P-period} gives the long-time periodic behaviour of the joint probability distribution $P_{\period}(x,c,t)$ over the fine timescale  in terms of the stationary marginal distribution $P_{\sts}(x)$ over the coarse-grained timescale. Therefore, Eqs.~\eqref{eq:sol-infty-free-protocol} and \eqref{eq:Ps-zero-tm-protocol}, inserted into Eq.~\eqref{eq:P-period}, also provide an analytical prediction for the long-time periodic joint probability distribution $P_{\period}(x,c,t)$.

\subsection{Numerical results}\label{sec:num-results}

Now we compare the analytical results stemming from our Markovian description of the feedback flashing ratchet with the numerical results obtained from the simulation of the system dynamics. In particular, we are interested in elucidating (a) the accuracy of the Markovian description we have developed in this paper, (b) the emergence of a long-time regime, as ensured by the $H$-theorem in Sec.~\ref{sec:H-theorem}, in which we expect the system to reach a time-periodic state, and (c) the specific results obtained in the limiting cases $\Delta t_m \to \infty$ and $\Delta t_m \to 0^+$, for which we have derived closed expressions. 

Units have been chosen as follows: $L$ as the length unit and $k_{B}T$ as the energy unit, i.e. we take $L=1$ and $k_{B}T=1$, whereas the time unit has been determined by taking $\gamma=1$---i.e. the time unit is $L^2/D=L^2\gamma /(k_B T)$. In general, we have employed a value of the force $F=5$ and several flashing protocols with different values of the uncertainty in the measurement of the position, specifically $\Delta x=0$, $\Delta x=0.5$, $\Delta x=0.75$, and $\Delta x=1$. We recall that $\Delta x=0$ corresponds to an error-free measurement, whereas $\Delta x=1$ corresponds to the open-loop protocol, since the uncertainty in the measurement extends to the whole interval due to the periodic boundary conditions. 

The theoretical results for the probability distributions have been obtained  with the procedure developed in Sec.~\ref{sec:formal-sol}, i.e. by concatenating Eqs.~\eqref{eq:evol-fp}, with the analytical expression for the Green function \eqref{eq:green-function} as an infinite series, and Eq.~\eqref{eq:evol-measure}. The specific expression of the Green function for the feedback flashing ratchet with the V-shaped potential is derived in~\ref{app:eigenvalues-and-eigenfunctions}, and we have truncated it after the $20$-th term for plotting the probability distributions. The simulation results have been obtained by combining the numerical integration of the Langevin equation~\eqref{eq:Langevin} with the Euler method between the control updates, and the corresponding update of the control variable at the times $t_k=k\Delta t_m$, as given by Eq. \eqref{eq:flashing-protocol}.  All the simulation results have been averaged over $10^6$ trajectories of the dynamics. 

We start by checking the accuracy of our Markovian description, along with the convergence to (i) a time-periodic state over the fine timescale and (ii) the tendency to a steady state over the coarse-grained timescale.

\begin{figure}
    \centering    \includegraphics[width=1\linewidth]{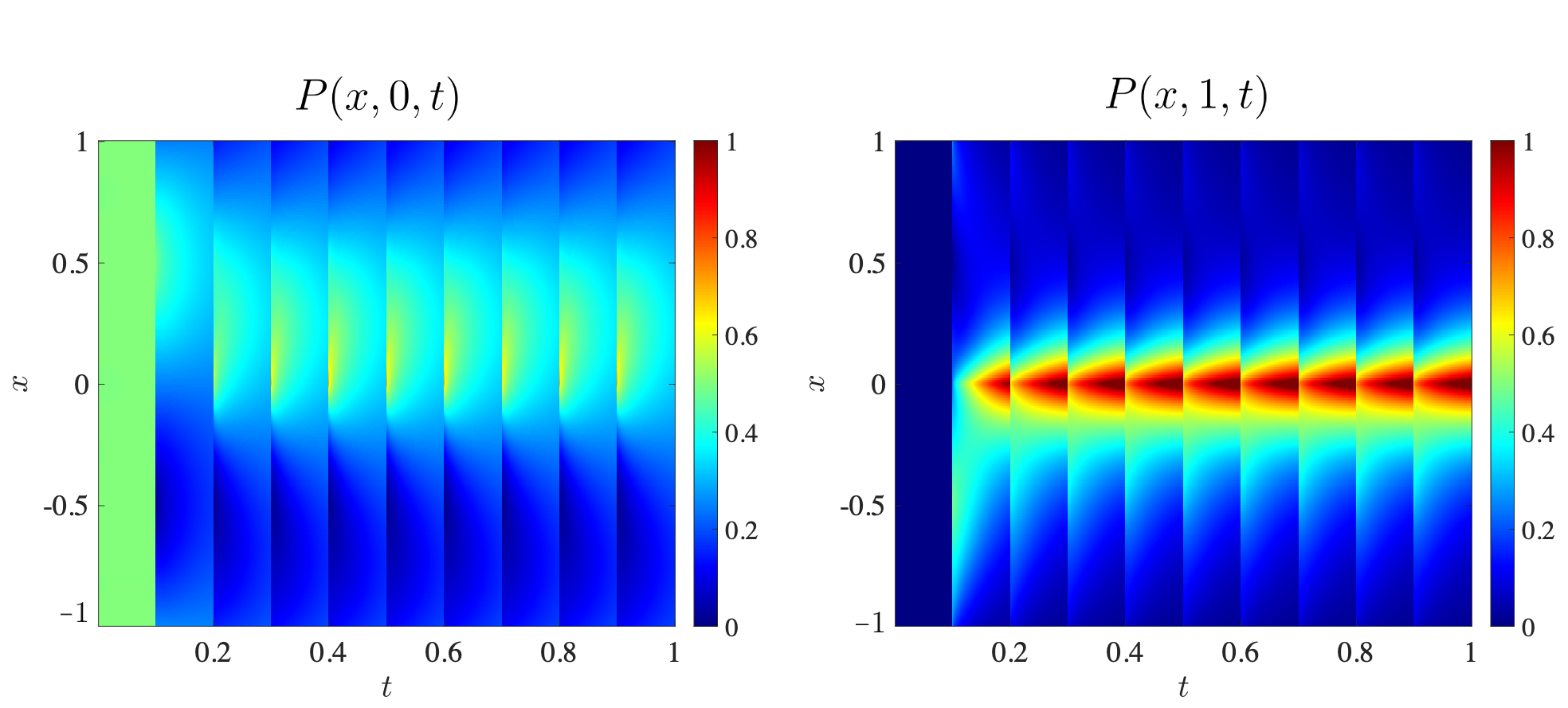}
    \caption{Analytical prediction for the time evolution of the joint probability distribution $P(x,c,t)$. Specifically, the plotted solution corresponds to parameters $\Delta t_m=0.1$,$F=5$ and $\Delta x=0.5$, in the time interval $[0,10 \Delta t_m]$. The tendency of the system to approach a time-periodic state is clearly observed. Despite the even parity of the V-shaped potential, $P(x,c,t)$ is not left-right symmetric due to the action of the feedback controller.}
    \label{fig:density_plot}
\end{figure}

In Fig.~\ref{fig:density_plot}, we present the analytical solution for the joint probability distribution of the position of the control $P(x,c,t)$. This analytical solution has been built for the initial condition in Eq.~\eqref{eq:initial-conditions}, in particular for a system with parameters $\Delta t_m=0.1$, $F=5$, and $\Delta x=0.5$, during a time interval in which there have been $10$ control updates at times $t_k=k\Delta t_m$, i.e. from $t=0$ to $t=10\Delta t_m$. Due to the specific initial conditions chosen, there is no time evolution in the first time interval $I_0=(0,\Delta t_m)$. Once the control is first updated at $t_1=\Delta t_m$, the system quickly evolves towards a time-periodic state  with period $\Delta t_m$ inherited from the periodicity of the control updates. This time-periodic state $P_{\period}(x,c,t)$ is the long-time behaviour expected from the $H$-theorem.

\begin{figure*}
  \centering
  \begin{subfigure}{1\textwidth}
    \centering
    \includegraphics[width=\linewidth]{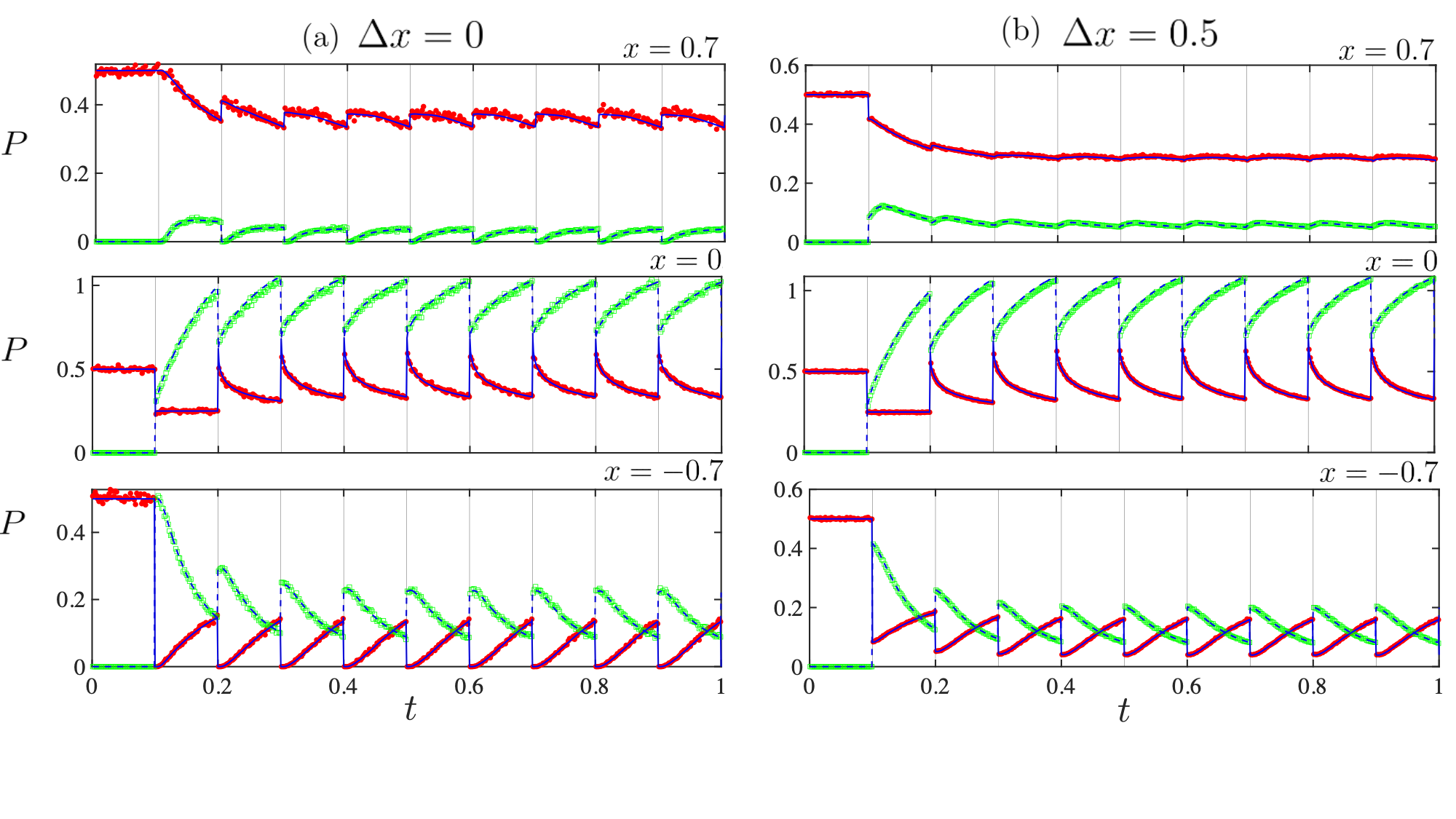}
  \end{subfigure}
  \begin{subfigure}{1\textwidth}
    \centering
    \includegraphics[width=\linewidth]{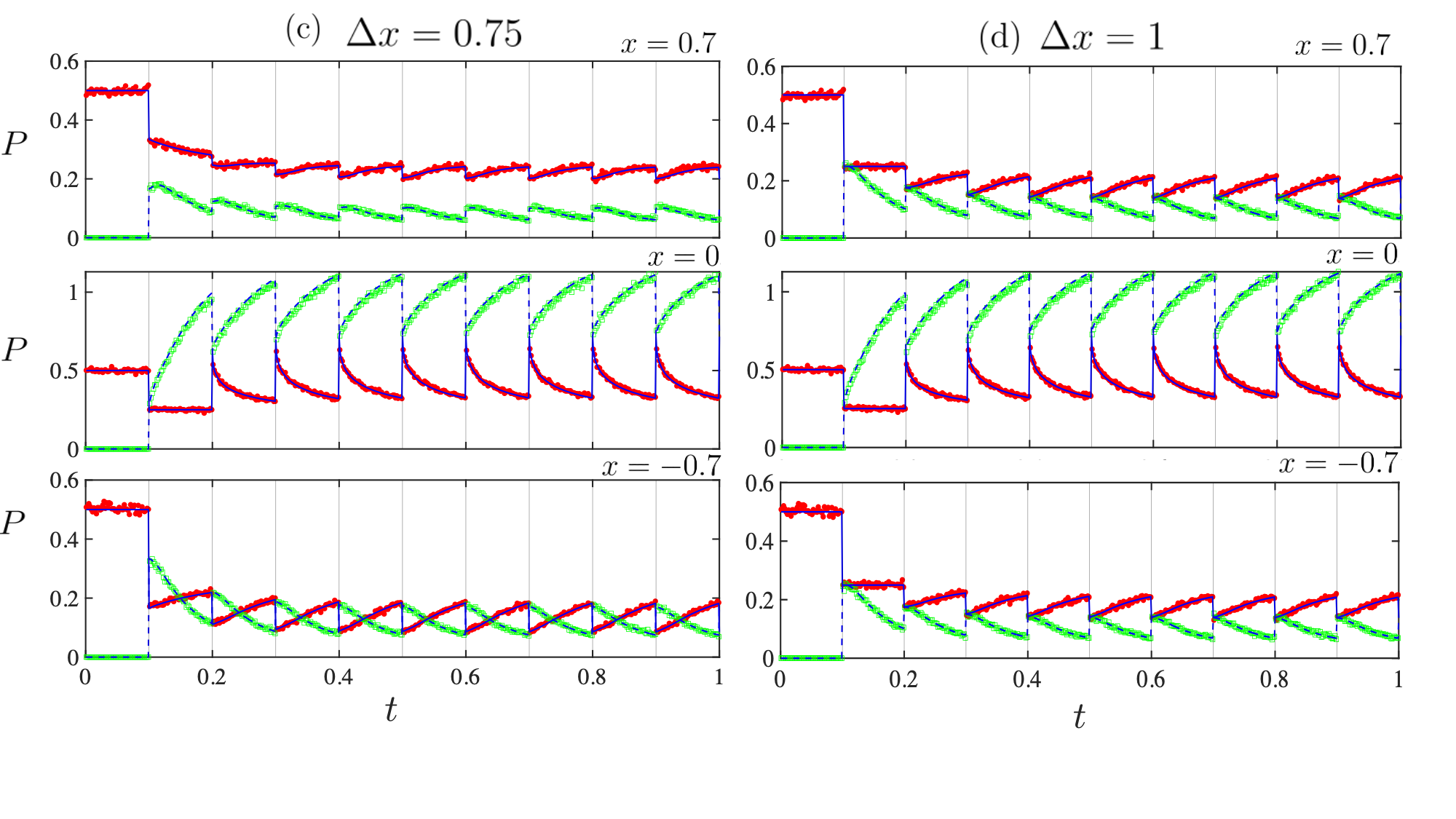}
  \end{subfigure}
   \caption{Time evolution of the joint probability $P(x,c,t)$. Four flashing protocols with different uncertainties have been considered in the different panels: (a) $\Delta x=0$, (b) $0.5$, (c) $0.75$, and (d) $1$. In each panel, we plot $P(x,c,t)$, with $c=0,1$, for three different values of $x$, $x=-0.7$ , $0$, and $0.7$. Numerical results (symbols) are compared with theoretical predictions (lines) for both $P(x,0,t)$ (red circles, solid line) $P(x,1,t)$ (green circles, dashed line). It is clearly observed the tendency of the system towards a time-periodic state with periodicity $\Delta t_m$ after a large enough number of control updates. Vertical lines represent the measurement times. The parameters used have been $\Delta t_m=0.1$ and $F=5$.}
    \label{fig:p_x_c}
\end{figure*}

In Fig.~\ref{fig:p_x_c}, we compare the analytical solution for the joint probability distribution $P(x,c,t)$ with the numerical integration of the dynamics. Specifically, we consider three different points in the spatial interval $[-1,1]$, $x=0$ and $x=\pm 0.7$, during the same time interval as in Fig.~\ref{fig:density_plot}. The times $t_k$ at which the control is updated are marked with the vertical lines. The agreement between the numerical results (symbols) and the theoretical prediction (lines) is excellent. Again, the time-periodic behaviour reached by the system after a large enough number of control updates $k$ is neatly observed. 

\begin{figure}
    \centering    \includegraphics[width=1\linewidth]{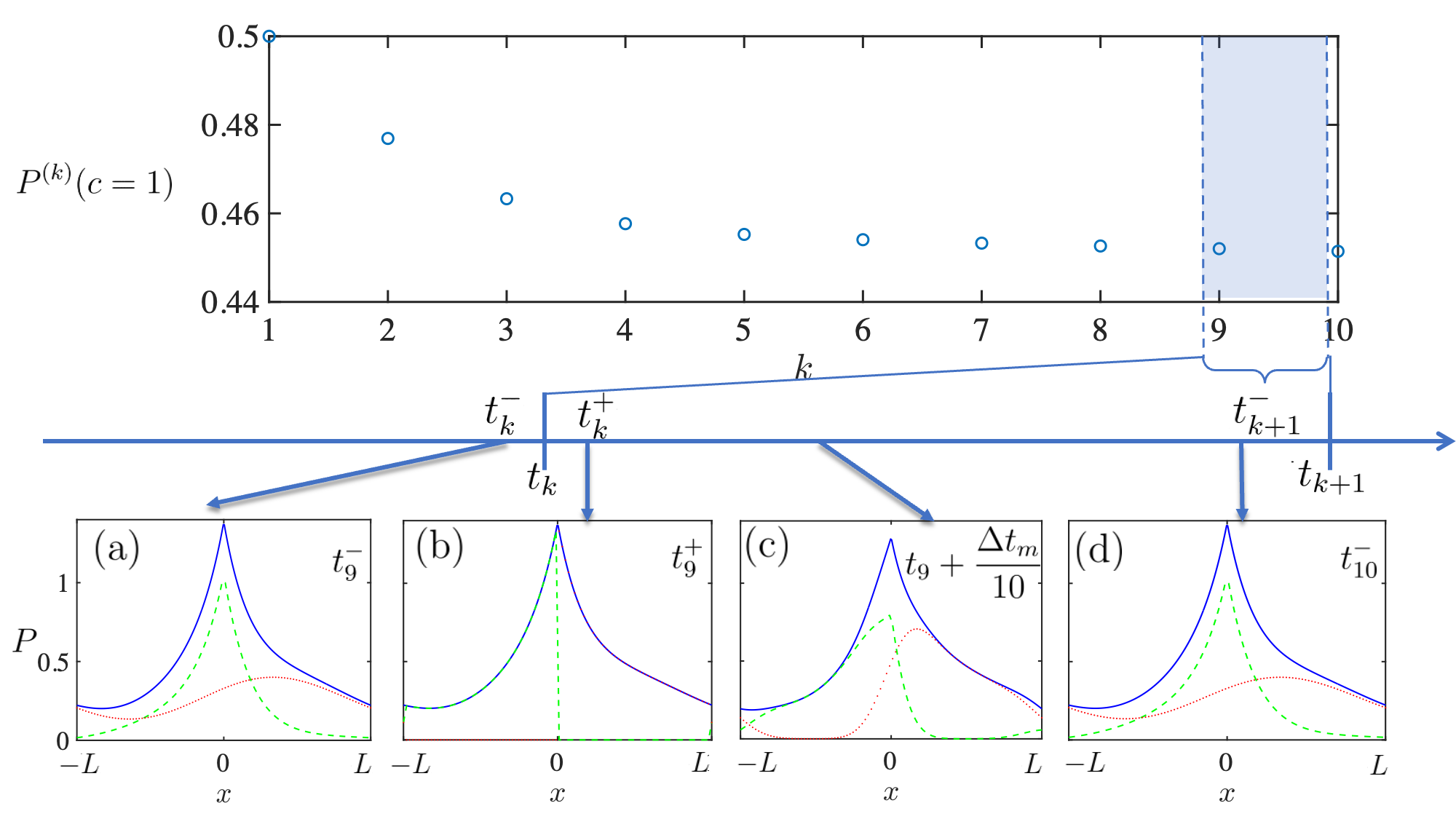}
    \caption{Compatibility of the evolution over the fine and coarse-grained timescales. System parameters are the same as in Fig.~\ref{fig:p_x_c}. The upper panel shows the evolution of $P^{(k)}(c)$ from $k=0$ to $k=10$, i.e. from $t=0$ to $t=10\Delta t_m$. The bottom panels focus on the evolution of  $P(x,t)$ (blue solid line), $P(x,0,t)$ (green dashed line) and $P(x,1,t)$ (red dotted line) in the last time interval $I_9=(t_9,t_{10})$ (shaded in the upper panel), i.e. between $t=9\Delta t_m$ to $t=10\Delta t_m$. Each panel collects the probabilities for different times, from (a) to (d): $t_9^-$, $t_9^+$, $t_9 +\Delta t_m/10$, and $t_{10}^-$. We appreciate how, if we just pay attention to the time instants $t_9^-$ and $t_9^+$ of panels (a) and (b), just before and after the measurement time, the marginal distribution $P(x,t)$ is continuous thereat, while the joint distribution $P(x,c,t)$ is not. Panel (c) shows a  snapshot of the time evolution of the probabilities in the time interval $I_9$, specifically at $t= t_9+\Delta t_m/10$. The steady state reached over the coarse-grained timescale is illustrated in panel (d), compare the profiles in panels (a) and (d).
    }
    \label{fig:evol_durig_I_k}
\end{figure}

Figure \ref{fig:evol_durig_I_k} illustrates the compatibility of the descriptions over the fine and coarse-grained timescales. We employ the same parameters as in Fig.~\ref{fig:p_x_c}. The top panel shows the evolution of $P(c=1,t)$, which only changes when the value of the control is updated at $t_k=k\Delta t_m$ after the measurement of the position at $t_k^-$.  After a few measurements, $P(c=1,t)$ reaches a steady state $P_{\sts}(c=1)$. In the bottom panels, we focus on the time evolution of $P(x,c,t)$ and $P(x,t)$ in one time interval $I_k=(t_k,t_{k+1})$ between control updates, in particular the last one, $I_9=(t_9,t_{10})$. At the end of the previous interval $I_8=(t_8,t_9)$, $P^{(k)}(x)\equiv P(x,t_k^-)$ has already reached the steady state $P_{\sts}(x)$  over the coarse-grained timescale for large enough $k$. However, over the finer timescale, both $P(x,c,t)$ and $P(x,t)$ evolve. When a measurement of the particle position takes place at the time $t_k^-$, the joint distribution changes abruptly from $P(x,c,t_k^-)$ to $P(x,c,t_k^+)$, following Eq. \eqref{eq:P-tk-+-v2}. This can also be seen in the bottom panels, where both $P(x,0,t)$ (green dashed line) and $P(x,1,t)$ (red dashed line) are shown.  However, $P(x,t_k^-)$ (blue solid line), is continuous at the control updates, i.e. $P(x,t_k^-)=P(x,t_k^+)$, as predicted by Eq. \eqref{eq:P-tk-+-marginal-continuity}---see panels (a) and (b). After the measurement, the joint probability $P(x,c,t)$ evolves from its initial state following its corresponding Fokker-Planck equation, i.e. under the action of the potential $V(x,c)=c\, V(x)$, and the marginal distribution $P(x,t)$ evolves too---see panel (c). At the final instant of the interval $I_9$, i.e. at $t_{10}^-$, $P(x,t)$ recovers the steady state shape $P_{\sts}(x)$ over the coarse-grained timescale---panel (d).

\begin{figure}
  \centering
  \begin{subfigure}{1\textwidth}
    \centering
    \includegraphics[width=\linewidth]{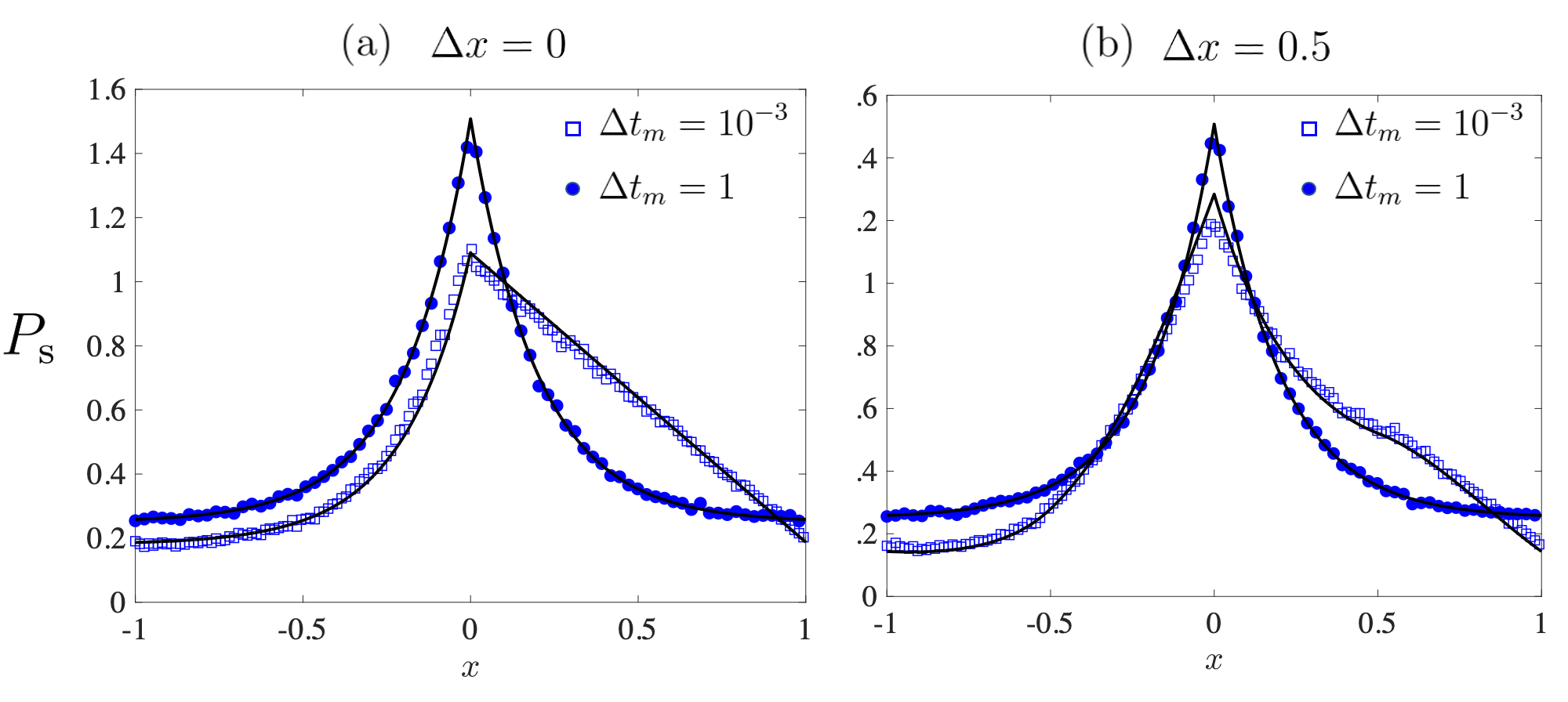}
  \end{subfigure}
  \hspace{1cm} 
  \begin{subfigure}{1\textwidth}
    \centering   \includegraphics[width=\linewidth]{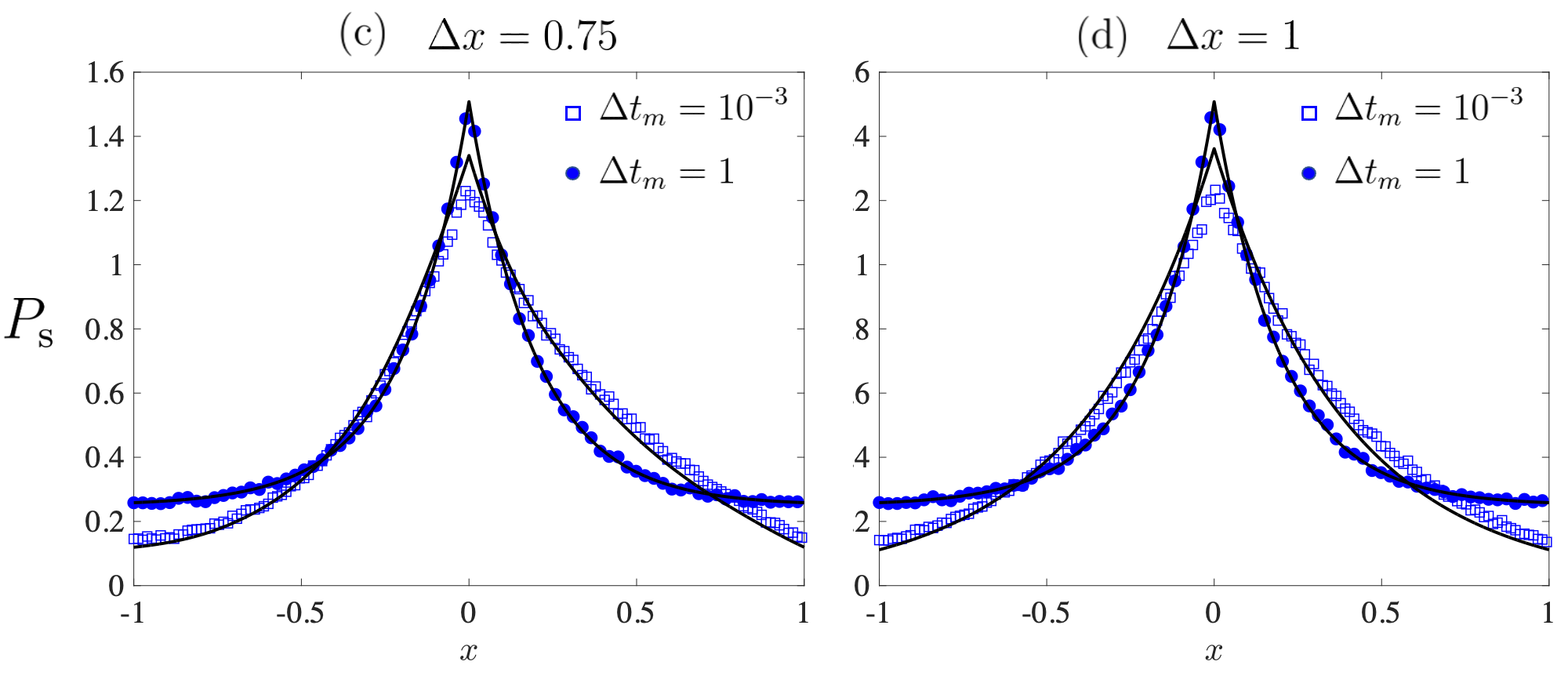}
  \end{subfigure}
\caption{Stationary marginal probability of the position over the coarge-grained timescale. In particular, we show the simulation result for $P(x,t_k^-)$ (symbols) for large enough $k$, and compare them with our analytical prediction for the stationary distribution $P_{\sts}(x)$ (lines). As in Fig.~\ref{fig:p_x_c}, panels (a)--(d) correspond to different flashing protocols, with increasing uncertainty from $\Delta x=0$ to $\Delta x=1$, for a force $F=5$. Here, two intervals between control updates are used: a short one $\Delta t_m=10^{-3}$ {\color{black}(open squares, solid line)}, and a long one $\Delta t_m=1$ {\color{black}(filled circles, solid line)}, to check the limits $\Delta t_m\to 0^+$ and $\Delta t_m\to\infty$, respectively.
   }
    \label{fig:limiting-cases}
\end{figure}

Now we proceed to inspect the specific results we have obtained in the limits $\Delta t_m \to \infty$ and $\Delta t_m \to 0^+$. Once more, we have employed the feedback flashing protocol~\eqref{eq:flashing-protocol}, for which we have analytical predictions for the steady distribution of the position over the coarse-grained timescale, as given by Eqs. \eqref{eq:sol-infty-free-protocol} ($\Delta t_m \to \infty$) and \eqref{eq:Ps-zero-tm-protocol} ($\Delta t_m \to 0^+$). In Fig.~\ref{fig:limiting-cases}, we have plotted the marginal probability $P^{(k)}(x)\equiv P(x,t_k)$, as defined in Eq. \eqref{eq:rho-k-def}, for large enough $k$, $k>k_r$.Note that ``enough" here depends on the time between control updates: $k_r$ decreases as $\Delta t_m$ increases, and $k_r\to 1$ for $\Delta t_m\to\infty$ because the steady state is reached after only one control update. In both cases, the numerical results for large $k$ (symbols) are very close to the analytical expressions for the steady distributions (lines). \textcolor{black}{We have numerically checked that the employed values of $\Delta t_m$ to simulate the limits are adequate: on the one hand, we have checked that $\Delta t_m=1$ is long enough to allow the system to reach the steady state at the end of each interval $I_k$ described in Sec.~\ref{sec:delta-tm-infty}; on the other hand, we have checked that $\Delta t_m = 10^{-3}$ is short enough for the system to follow the effective dynamics described in Sec.~\ref{sec:delta-tm-zero}.} 

\begin{figure}
    \centering\includegraphics[width=0.75\linewidth]{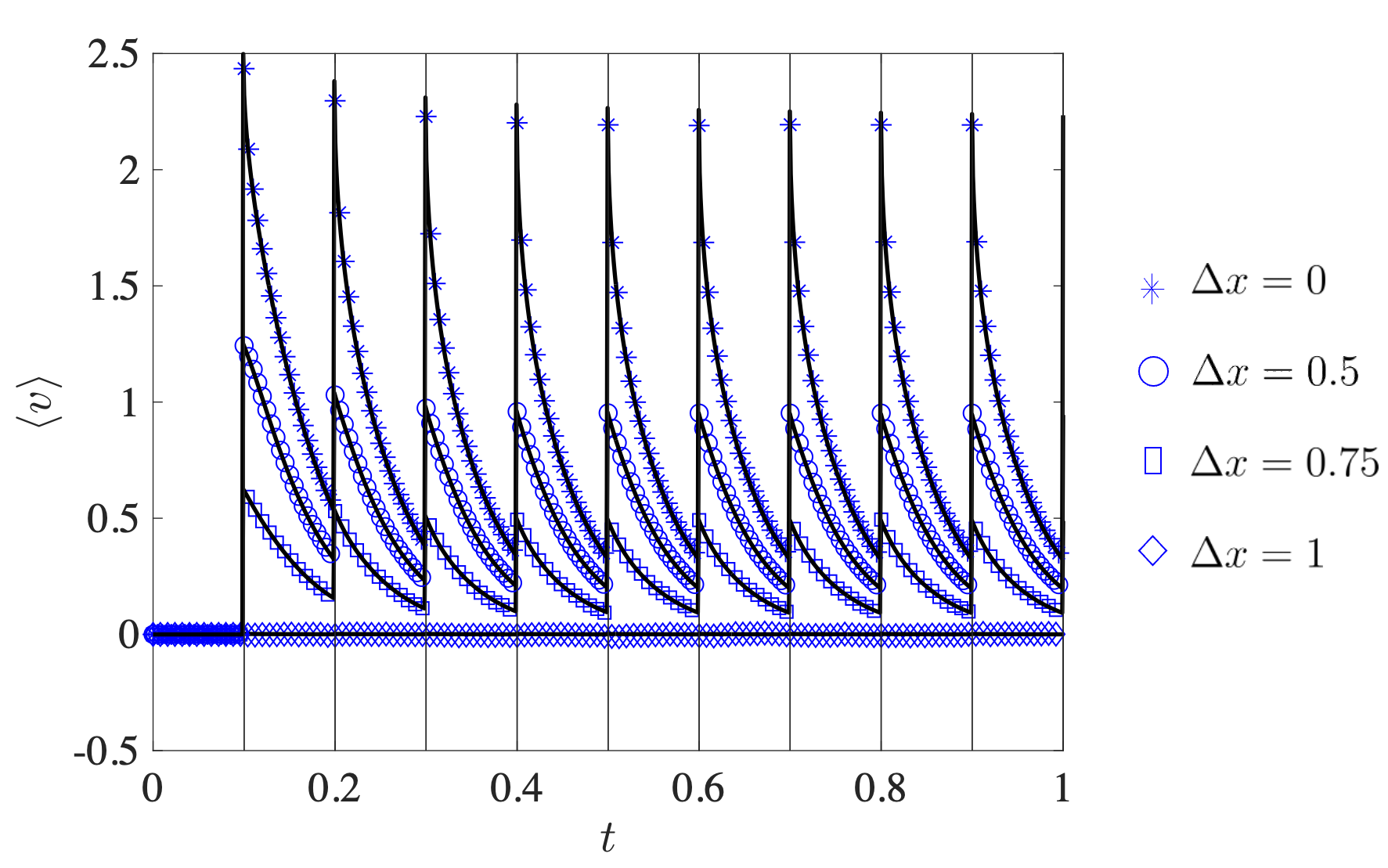}
    \caption{Average velocity $\expval{v}$ of the particle as a function of time. System parameters are the same as in Fig.~\ref{fig:p_x_c}. Simulation (symbols) and theoretical (lines) results  have been plotted for different values of the uncertainty, as usual $\Delta x=0$ (stars), $0.5$ (circles), $0.75$ (squares), and $1$ (diamonds). As $\Delta x$ increases, $\expval{v}$ decreases from its maximum value for $\Delta x=0$ (error-free feedback) to zero for $\Delta x=1$ (open-loop). }
    \label{fig:velocity}
\end{figure}

Finally, we present results for the average velocity of the particle. With periodic boundary conditions, the average velocity of the particle is calculated by considering the average of the velocity field $v(x,c,t)\equiv J(x,c,t)/P(x,c,t)$---see \ref{app:periodic-bc}. For our specific case, we have
\begin{align}\label{eq:velocity}
    \expval{v}(t)&\equiv \sum_c\int_{-L}^{L}dx\, v(x,c,t) P(x,c,t)=\sum_c \int_{-L}^{L}dx \, J(x,c,t) \nonumber \\
    &=-\gamma^{-1}\sum_c\int_{-L}^{L}dx\,
    V'(x,c)P(x,c,t)=\gamma^{-1}F \int_{-L}^{L}dx\,
    \sgn(-x) P(x,1,t). 
\end{align}
Our family of flashing protocols depends on the uncertainty $\Delta x$ in the position measurement, varying from error-free feedback when $\Delta x=0$ to open-loop control when $\Delta x=1$. Figure \ref{fig:velocity} presents $\expval{v}$ as a function of time, for different values of $\Delta x$; $\expval{v}$ decreases with $\Delta x$, vanishing for $\Delta x=1$. This trend can be understood as follows: the physical mechanism behind the creation of a directed flux is the breaking of the spatial symmetry. Since our potential is left-right symmetric, the symmetry breaking must come from the feedback control, which uses information from the position to update the potential state. For $\Delta x=1$, we have an open-loop protocol that no longer gathers spatial information from the system and, therefore, no flux is created.

It is also interesting to link the {\color{black}mutual information between the particle position and the controller variable,} $I(X,C)$, and the mean velocity of the particle just after the measurement times, i.e at $t=t_k^+$, to shed light on the role played by the mutual information on the flow. 
The average velocity just after the measurement is given by
\begin{equation}
    \av{v}(t_k^+)=\gamma^{-1}F \int_{-L}^{L}  dx \, \sgn(-x) \Theta(1|x) P_{\sts}(x) ,
\end{equation}
particularising Eq. \eqref{eq:velocity} to $t=t_k^+$, for large enough $k$.  The maximum average velocity of the periodic state reached in the long time limit is given by $\av{v} (t_k^+)$, as Fig. \ref{fig:velocity} shows.

First, we focus on the $\Delta t_m \to \infty$ case, for which the mutual information is given by Eq.~\eqref{eq:Is-def}, with $P_{\sts}(c)$ and $P_{\sts}(x)$ being given by  Eqs.~\eqref{pc_infty} and \eqref{eq:sol-infty-free-protocol} of Sec.~\ref{analytical-results}. 
For the average velocity, it is advantageous to take into account that $\Theta(1|x)-1/2$ is an odd function of $x$---see Fig. \ref{fig:protocols}, so we can write
\begin{align}
      \av{v}(t_k^+)&=\frac{\gamma^{-1}F}{2} \int_{-L}^{L}  dx \,\sgn(-x) \Theta(1|x) \left[P_{\sts}^{(0)}(x)+P_{\sts}^{(1)}(x)\right] \nonumber \\
      &=\frac{\gamma^{-1}F}{2} \int_{-L}^{L}  dx \,\sgn(-x) \left[\Theta(1|x)-\frac{1}{2}\right]\left[P_{\sts}^{(0)}(x)+P_{\sts}^{(1)}(x)\right] \nonumber \\ & =\gamma^{-1}F \int_{0}^{L}  dx \,\sgn(-x) \left[\Theta(1|x)-\frac{1}{2}\right]\left[P_{\sts}^{(0)}(x)+P_{\sts}^{(1)}(x)\right] \nonumber \\
      &=\frac{\gamma^{-1}F}{2}-\gamma^{-1} F \int_0^L dx \, \Theta(1|x) \left[P_{\sts}^{(0)}(x)+P_{\sts}^{(1)}(x)\right].
\label{eq:velocity-tm-infty}
\end{align}
We have repeatedly used that both $P_{\sts}^{(0)}$ and $P_{\sts}^{(1)}$ are even functions of $x$. The second term of Eq.~\eqref{eq:velocity-tm-infty} is  non-positive. Therefore, $\av{v} (t_k^+)$ reaches a maximum value when this term vanishes for perfect closed-loop control, whereas $\av{v} (t_k^+)$  vanishes for open-loop control, for which $\Theta(1|x)=1/2$. This is consistent with the fact that, for open-loop protocols and a symmetric potential, there is no symmetry breaking that favours a particular direction and creates a directed flux.

Second, for the case $\Delta t_m \to 0^+$, the mutual information is again provided by Eq.~\eqref{eq:Is-def}, but now $P_{\sts}(x)$ is given by Eq.~\eqref{eq:Ps-zero-tm-protocol} and $P_{\sts}(c)=\int_{-L}^L dx\, \Theta(c|x) P_{\sts}(x)$, as given by Eq.~\eqref{eq:P-k-C-sts}. For the flux,  the average velocity equals that of a particle moving under the effective potential $V_{\eff}(x)$. Taking into account that the current is homogeneous in the steady state, 
\begin{align}\label{eq:velocity-tm-0}
     \av{v}(t_k^+) &= 2L J_{\sts}^{\eff}=K \,\frac{2L}{\beta\gamma} \, \frac{1- e^{-\beta [V_{\eff}(-L)-V_{\eff}(L)]}}
  {\int_{-L}^{L}dx\, e^{\beta V_{\eff}(x) } }, 
\end{align}
where  $K>0$ is the normalisation constant of the probability in Eq.~\eqref{eq:Ps-zero-tm-protocol} and $J_{\sts}^{\eff}$ is the steady current for the effective potential---see~\ref{app:stationary-sol}, specifically Eq.~\eqref{eq:J-sol}.
Again,  $\av{v} (t_k^+)$ reaches its maximum value for error-free closed-loop control, which corresponds to the maximum difference $V_{\eff}(-L)-V_{\eff}(L)$, whereas $\av{v} (t_k^+)$  vanishes for open-loop control, for which $V_{\eff}(x)$ becomes an even function of $x$---see Fig.~\ref{fig:effective_potential}.

\begin{figure}
    \centering
    \includegraphics[width=0.75\linewidth]{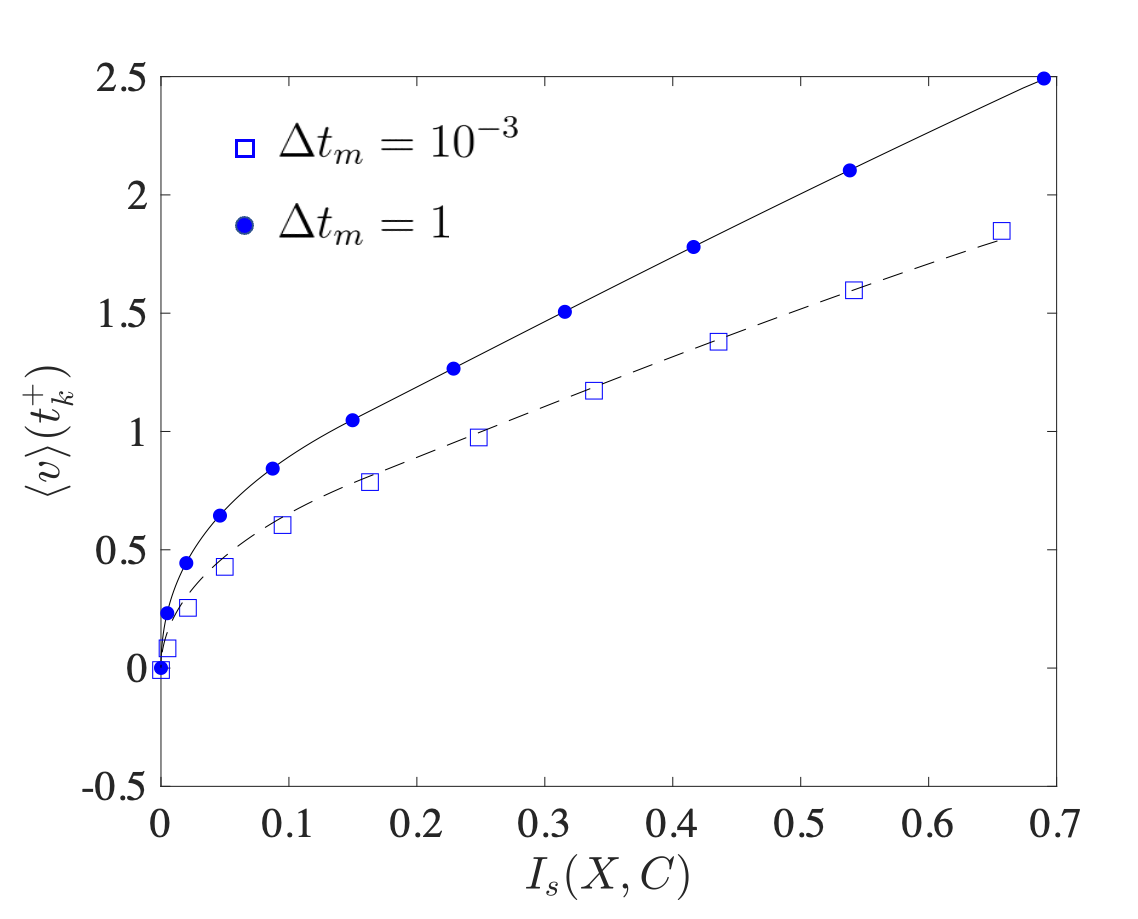}
    \caption{
   \textcolor{black}  {Average velocity versus the mutual information between the particle's position and the control state. Specifically, both quantities have been evaluated just after the measurement, i.e. at $t_k^+$, in the long-time regime reached for large enough $k$, for different values of $\Delta x$. As in the previous figures, $F=5$. Theoretical  (lines)  and simulation (symbols) data are plotted for the limiting cases $\Delta t_m \to 0^+$ (dashed line, open squares) and $\Delta t_m \to \infty$ (solid line, filled circles).  From left to right, the plotted data correspond to an uncertainty in the measurement decreasing from $\Delta x=1$ (open-loop) to $0$ (error-free closed-loop). In agreement with the physical intuition, both the flux and the mutual information increase from left to right, from zero to their maximum values.} 
}
    \label{fig:info_vs_velocity}
\end{figure}

{Figure \ref{fig:info_vs_velocity} shows a parametric plot of $\av{v} (t_k^+)$ versus $I_{\sts}(X,C)$, obtained by varying the uncertainty in the measurement $\Delta x$, for both limits, $\Delta t_m \to 0^+$ (left panel) and $\Delta t_m \to \infty$ (right panel). We recall that, by varying $\Delta x$, we go from the error-free closed-loop protocol ($\Delta x=0$) to the open-loop protocol ($\Delta x=L$). The monotonic behaviour of the velocity as a function of the {\color{black} mutual information} reinforces the intuitive picture of the system: {\color{black} the larger the mutual information after the measurement,} the larger the flow, which varies from its minimum value when $\Delta x=L$ (open-loop protocol) to its maximum value when $\Delta x=0$ (error-free closed-loop protocol). For $\Delta x=L$, the protocol becomes $\Theta(c|\cancel{x})=\Theta(c)=1/2$, or, what is the same, the control switches on and off the potential randomly, with probability $1/2$. In this case, the mutual information between the control state and the particle position is zero since they become independent variables: $P_{\sts}(c)=\Theta(c)$ in Eq.~\eqref{eq:Is-def}. Conversely, for $\Delta x=0$, the control becomes a deterministic function of the position and the second term on the rhs of Eq.~\eqref{eq:Is-def} vanishes,  since $\Theta(0|x)=\eta(x)$ and $\Theta(1|x)=1-\eta(x)=\eta(-x)$. Therefore, the mutual information between the control and the particle's position varies from its maximum value for error-free feedback, $\Delta x=0$, to zero for open-loop control, $\Delta x=L$. This is in complete analogy to the behaviour for the flux $\av{v} (t_k^+)$ as a function of $\Delta x$ stemming from Eqs.~\eqref{eq:velocity-tm-infty} and \eqref{eq:velocity-tm-0}. }

\textcolor{black}{In Fig.~\ref{fig:info_vs_velocity}, it is observed that $\av{v(t_k^+)}$ is lower for $\Delta t_m \to 0^+$ than for $\Delta t_m \to \infty$, which may seem to contradict the physical intuition at first sight---one expects a larger flux for $\Delta t_m\to 0^+$. In this respect, it must be stressed that the physically relevant quantity would be the time-averaged velocity over a period,
\begin{equation}\label{eq:time-aver-flux}
    \overline{v}_{\sts} \equiv \frac{1}{\Delta t_m}\int_{t_k^+}^{t_{k+1}^-} dt\, \av{v(t)},
\end{equation} 
for large enough $k$. It is indeed $\overline{v}_{\sts}$ that has the physically expected behaviour. For $\Delta t_m\to 0^+$, $\overline{v}_{\sts}$ coincides with $\av{v}(t_k^+)$. For $\Delta t_m\to\infty$, $\av{v(t)}$ tends to zero at the end of the interval between measurements, at $t=t_{k+1}^-$. This makes the integral $\int_{t_k^+}^{t_{k+1}^-} dt\, \av{v(t)}$ in Eq.~\eqref{eq:time-aver-flux} go to a well-defined limit as $\Delta t_m\to\infty$ and, as a consequence, $\overline{v}_{\sts}$ vanishes.}

\section{Summary and conclusions}
\label{sec:conclusions}

We have considered a wide class of feedback controlled systems, in which an overdamped particle moves in a fluctuating potential that is changed by an external controller depending on the instantaneous position of the particle. The controller acts regularly at times $t_k=k\Delta t_m$, $k=1,2,\ldots$, which naturally divides the time evolution into intervals $I_k=(t_k,t_{k+1})$. In each trajectory of the dynamics, the controller measures the position of the particle $x_k^-\equiv x(t_k^-)$ just before $t_k$, and the potential felt by the particle at time $t_k^+$ (and during the whole interval $I_k$) is $V(x,c_k)$ with probability given by a generic function $\Theta(c_k|x_k^-)$. This allows for the inclusion of imperfect measurements with a certain uncertainty $\Delta x$. As a consequence, our framework includes as limiting cases error-free feedback control, where $\Theta(c|x)$ depends on $x$ but equals either $0$ or $1$ for any $x$, and open-loop, i.e. no actual feedback, control, where $\Theta(c|x)$ is independent of $x$.

In general, neither the position $X$ nor the controller variable $C$ is a Markov process. However, the joint stochastic process $(X,C)$ is Markovian, even for imperfect control. The joint probability distribution $P(x,c,t)$ evolves following a differential Chapman-Kolmogorov equation, which has two contributions: (i) a Fokker-Planck contribution, which accounts for the time evolution under the potential $V(x,c)$ between control updates, i.e. in the time intervals $I_k$, and (ii) a master equation contribution, which accounts for the change of the potential at the times of the control updates $t_k$. This general Markovian framework allows us to give a formal analytical solution for the time evolution of the joint probability distribution, in terms of an expansion in the eigenmodes and eigenfunctions of the potentials $V(x,c)$, for an arbitrary frequency of the control updates, i.e. for arbitrary $\Delta t_m$.

We have proved an $H$-theorem for the joint stochastic process $(X,C)$, by constructing a Lyapunov functional $H(t)$ that is the Kullback-Leibler divergence between two solutions of the differential Chapman-Kolmogorov equation. We have shown that $H(t)$ is monotonically decreasing, both during the Fokker-Planck evolution in the time intervals $I_k$ and at the control updates at the times $t_k$ delimiting them. This $H$-theorem ensures that the joint probability distribution $P(x,c,t)$ reaches a long-time regime, independent of the initial conditions. On a physical basis, the long-time behaviour of the joint distribution, $P_{\period}(x,c,t)$, is expected to be time-periodic with period $\Delta t_m$, inherited from the control.

The above Markovian description holds over the (fine) timescale that includes both the times $t_k$ at which the control is updated and the time intervals $I_k$ between control updates. One may go to a coarse-grained timescale that only considers the times $t_k^+$ just after the control updates. Even for imperfect measurements, the control is enslaved to the position at these times and thus the position $X$ alone becomes Markovian. Over this coarse-grained timescale, the probability distribution of the position approaches a stationary distribution $P_{\sts}(x)$. Interestingly, finding this stationary distribution over the coarse-grained timescale makes it possible to derive the time-periodic distribution $P_{\period}(x,c,t)$ over the fine timescale, which the system approaches for long times as a consequence of the $H$-theorem. {The mutual information between the particle's position and the control state at the times $t_k^+$ can also be calculated from our framework. Moreover,} in the limits of very frequent ($\Delta t_m\to 0^+$) or very infrequent ($\Delta t_m\to\infty$) control updates, explicit expressions for the stationary distribution $P_{\sts}(x)$---and thus for $P_{\period}(x,c,t)$---have been derived.

The just described theoretical framework has been applied to a specific model:  the feedback flashing ratchet. Therein, we have compared the theoretical expressions obtained with numerical simulations of the system dynamics---the latter consisting in the concatenation of the numerical integration of the Langevin equation in the time intervals $I_k$ with the control updates at the times $t_k$. The agreement between the numerics and the theoretical results has been excellent in all cases. { Among other things, our study shows how the average flux increases with the mutual information between the particle and the controller variables just after the measurement. In this line, an interesting prospect is to derive upper bounds on the current by using inequalities arising from fluctuation theorems~\cite{sagawa_fluctuation_2012} or thermodynamic uncertainty relations~\cite{potts_thermodynamic_2019}. }

The impact of the information gathered by a feedback controller in the thermodynamic balance, since it reduces the entropy of the particle, is an active line of research~\cite{cao_thermodynamics_2009,sagawa_minimal_2009,sagawa_thermodynamics_2012,abreu_thermodynamics_2012,sagawa_nonequilibrium_2012,deffner_information_2013,van_vu_uncertainty_2020,ruiz-pino_information_2023}. For the particular situation of the feedback flashing ratchet, this has recently been done~\cite{ruiz-pino_information_2023} by considering the full history of control actions $(c_1,c_2,\ldots)$---due to the non-Markovianity of the position. The Markovian description for the joint process $(X,C)$ opens new perspectives in this direction, since one may consider the thermodynamic balance for the Shannon entropy of the joint process,
\begin{equation}\label{eq:Shannon-entropy}
    S=-\sum_c \int dx\, P(x,c,t) \ln P(x,c,t).
\end{equation}
We would like to stress that the connection between this Shannon entropy and the entropy of the full history of control actions is not trivial, and deserves to be 
further investigated. \textcolor{black}{In this regard, it seems relevant to look into the relation between the entropy reduction calculated from the full history of control actions~\cite{ruiz-pino_information_2023} and the mutual information between the position and the control variable in the Markovian picture developed here. Also, it is also worth looking for bounds on the flux as a function of the mutual information~\cite{cao_information_2009}.}

The use of the Shannon entropy~\eqref{eq:Shannon-entropy} for the Markovian joint process $(X,C)$ opens other interesting perspectives for future research. { For example, it seems worth looking for the transposition to our Markovian framework of the splitting of the entropy production into ``housekeeping'' and ``excess'' contributions~\cite{sekimoto_langevin_1998,hatano_steady-state_2001,speck_integral_2005,van_den_broeck_three_2010,seifert_stochastic_2012}. This is relevant for deriving the tightest bound on the work extracted from the feedback-controlled system and therefrom the most accurate  definition of the efficiency for information engines, which involves the incorporation of the mutual information to the thermodynamic balance~\cite{cao_thermodynamics_2009,sagawa_minimal_2009,sagawa_thermodynamics_2012,abreu_thermodynamics_2012,sagawa_nonequilibrium_2012,deffner_information_2013,van_vu_uncertainty_2020,ruiz-pino_information_2023}. Also, recent developments of information geometry ideas, which propose alternative splittings of the entropy production~\cite{dechant_minimum_2022,dechant_geometric_2022_PRE,dechant_geometric_2022_PRR,kamijima_thermodynamic_2023,ito_geometric_2024,yoshimura_housekeeping_2023,yoshimura_two_2024}, could give better refinements of the second principle in this context and thus improve our current understanding of the second principle in systems with feedback control. }

\section*{Acknowledgments}
We acknowledge financial support from Grant PID2021-122588NB-I00 funded by MCIN/AEI/10.13039/501100011033/ and by ``ERDF A way of making Europe'', and from Grant ProyExcel\_00796 funded by Junta de Andalucía's PAIDI 2020 programme. NRP acknowledges support from the FPU programme  through Grant FPU2021/01764. We also thank C.~A.~Plata and F.~J.~Cao for really useful discussions.

\section*{Data availability}
The codes employed for generating the data that support the findings of this study, together with the Mathematica notebooks employed for producing the figures presented in the paper, are openly available in the~\href{https://github.com/fine-group-us/markovian-description-of-feedback-ratchets}{GitHub page} of University of Sevilla's FINE research group.

\appendix

\section{Periodic boundary conditions}\label{app:periodic-bc}
In our system, the particle is moving under a periodic force with
spatial period $2L$. We can thus study the system on the whole real
line, $-\infty<x<+\infty$, or only in one period of the potential,
e.g. $-L\le x\le L$. In this section, we would like to clarify the
relation between the two descriptions. 

Our starting point is the Fokker-Planck equation~\eqref{eq:FP-Jk} in the time intervals $I_k$, i.e.
\begin{align}
  \label{eq:FP-continuity}
  \partial_{t}P(x,c,t)&=-\partial_{x}J(x,c,t),  &&J(x,c,t) =-\gamma^{-1}\left[\tilde{V}'(x,c)P(x,c,t)+k_{B}T\,\partial_{x}P(x,c,t)
  \right],
\end{align}
The force is periodic, so that $\tilde{V}'(x)=\tilde{V}'(x+2L)$, and therefore
$P(x+2nL,c,t)$, with $n\in\mathbb{Z}$, is also a solution of the Fokker-Planck
equation. Then, we can construct the
probability of $x$ modulo $L$,
\begin{equation}
  \label{eq:P-mod-L}
  P_{\modulo}(x,c,t)\equiv \sum_{n}P(x+2nL,c,t), \quad \abs{x}\le L.
\end{equation}
i.e. we treat all the points that differ by $2nL$ as equivalent and
thus sum all their probabilities.

When constructing $P_{\modulo}(x,t)$, we restrict the problem
to one period $[-L,L]$ and then we expect this to be equivalent to considering periodic boundary conditions. And, in fact, this is basically
what is being done, but let us understand it on a solid basis. Equation~\eqref{eq:P-mod-L} entails that $P_{\modulo}(-L,c,t)=P_{\modulo}(L,c,t)$. Moreover, we have
\begin{equation}
  \label{eq:norm-conserv}
  0=\frac{d}{dt}\int_{-\infty}^{+\infty}dx P(x,c,t)=
  \frac{d}{dt}\int_{-L}^{L}dx
  P_{\modulo}(x,c,t)=J_{\modulo}(-L,c,t)-J_{\modulo}(L,c,t),
\end{equation}
where
\begin{equation}
  \label{eq:J-mod-L}
  J_{\modulo}(x,c,t)\equiv\sum_{n}J(x+2nL,c,t)=-\gamma^{-1}
  \left[\tilde{V}'(x,c)P_{\modulo}(x,c,t)+k_{B}T\,\partial_{x}P_{\modulo}(x,c,t)
  \right], 
\end{equation}
for $\abs{x}\le L$. Therefore, the boundary conditions for the problem in $[-L,L]$ are
\begin{equation}
  \label{eq:bc-0-L}
  P_{\modulo}(-L,c,t)=P_{\modulo}(L,c,t), \quad
  J_{\modulo}(-L,c,t)=J_{\modulo}(L,c,t). 
\end{equation}

\subsection{Particle velocity in the periodic picture}\label{app:particle-velocity}

It is also interesting to link the average velocity of the particle on the whole real line with the average flux in the periodic picture. The evolution equation for the average position on
the whole real line is readily derived from the Fokker-Planck equation:
\begin{equation}
  \label{eq:evol-eq-av-x}
  \av{v}\equiv\frac{d}{dt}\sum_c\int_{-\infty}^{+\infty} dx \,x P(x,c,t)=
  -\frac{1}{\gamma}\sum_c\int_{-\infty}^{+\infty}dx\, \tilde{V}'(x,c)P(x,c,t)=-\frac{1}{\gamma}\av{\tilde{V}'(x,c)},
\end{equation}
where we have integrated by parts. When one goes to the interval $[-L,L]$ with periodic boundary conditions, $\av{v}\ne d/dt \sum_c\int_{-L}^{L} dx\, x P_{\modulo}(x,c,t)$. Where is the information of the average velocity stored? The answer is that it is stored in the spatial average of the current:
\begin{align}
  \sum_c\int_{-L}^{L}dx J_{\modulo}(x,c,t)=&-\frac{1}{\gamma}\sum_c\int_{-L}^{L}dx\,
\tilde{V}'(x,c)P_{\modulo}(x,c,t) \nonumber \\
&-\frac{k_{B}T}{\gamma}\sum_c\cancelto{0}
{\left[P_{\modulo}(L,c,t)-P_{\modulo}(-L,c,t)\right]}.
\end{align}
Taking into account that $\tilde{V}'(x,c)$ is spatially periodic,
\begin{align}
    \sum_c\int_{-L}^{L}dx J_{\modulo}(x,c,t) & =
-\frac{1}{\gamma}\sum_c\int_{-\infty}^{+\infty}dx\, \tilde{V}'(x,c)P(x,c,t)=
-\frac{1}{\gamma}\av{\tilde{V}'(x,c)}=\av{v} ,
   \label{eq:current-av-veloc}
\end{align}
which completes the proof.

\section{Stationary distributions}\label{app:stationary-sol}

Let us consider the stationary state for the Fokker-Planck equation with force $-\tilde{V}'(x)$ and periodic boundary conditions---consistently with our notation in the main text, we omit the subindex ``$\modulo$'' from now on. At the steady state, the current is uniform in the whole system:
\begin{equation}
  \label{eq:Js}
  J_{\sts}=-\gamma^{-1}\left[\tilde{V}'(x)P_{\sts}(x)+
    k_{B}T\,\partial_{x}P_{\sts}(x)\right] .
\end{equation}
Integrating this equation, we get
\begin{equation}
  \label{eq:Ps-sol}
  P_{\sts}(x)=K e^{-\beta \tilde{V}(x)}-\beta\gamma J_{\sts}\int_{-L}^{x}dx'
  e^{-\beta\left[\tilde{V}(x)-\tilde{V}(x')\right] }  ,
\end{equation}
where $K$ is a normalisation constant. By imposing the periodicity of $P$, $P_{\sts}(-L)=P_{\sts}(L)$, we get 
\begin{equation}
  \label{eq:periodic-P}
  K e^{-\beta \tilde{V}(-L)}=K e^{-\beta \tilde{V}(L)}-\beta\gamma J_{\sts}\int_{-L}^{L}dx
  e^{-\beta\left[\tilde{V}(L)-\tilde{V}(x)\right] } .
\end{equation}
This equation tells us that $J_{\sts}\ne 0$ if and only if $\tilde{V}(-L)\ne
\tilde{V}(L)$. Solving for $J_{\sts}$,
\begin{equation}
  \label{eq:J-sol}
  \beta\gamma J_{\sts}=K \frac{e^{-\beta \tilde{V}(L)}- e^{-\beta \tilde{V}(-L)}}
  {\int_{-L}^{L}dx\,
  e^{-\beta\left[\tilde{V}(L)-\tilde{V}(x)\right] } }, 
\end{equation}
and one has finally that
\begin{equation}
  \label{eq:Ps-sol-v2}
  P_{\sts}(x)=K \left[e^{-\beta \tilde{V}(x)}- \frac{e^{-\beta \tilde{V}(L)}-e^{-\beta
        \tilde{V}(-L)}}{\int_{-L}^{L}dx'
  e^{-\beta\left[\tilde{V}(L)-\tilde{V}(x')\right] }  }\int_{-L}^{x}dx'
  e^{-\beta\left[\tilde{V}(x)-\tilde{V}(x')\right] }  \right] .
\end{equation}

Note that $K>0$ in order to have a meaningful stationary distribution,
because $P_{\sts}(-L)=K e^{-\beta \tilde{V}(-L)}$. This implies that $J_{\sts}>0$ if
$\tilde{V}(L)<\tilde{V}(-L)$, in agreement with the physical intuition. The sign of $J_{\sts}$
is reversed if the potential is reflected through the midpoint of the
interval, i.e. if we have a potential $\tilde{V}_2(x)=\tilde{V}(-x)$. In that
case, it is easy to see that
$P_2(x,t)=P(-x,t)$ is a solution of the
Fokker-Planck equation with force $-{\tilde{V}_2}'(x)$. In the steady
state, $P_2^{\sts}(x)=P_{\sts}(-x)$ and
thus $J_{2,\sts}=-J_{\sts}$.

\subsection{Conditional steady distributions}\label{app:cond-steady-distr}
  
Let us consider the time evolution of the marginal distribution for
the control $P(c,t)$, which we obtain by integrating
\eqref{eq:diff-CK} over $x$:
\begin{align}
  \partial_{t}P(c,t)=\!\!
  \sum_{k}\delta(t-t_{k}^{-})\!\! \int \!\!dx \Big[ \Theta(c|x)\sum_{c'\ne
c}P(x,c',t)-\left[1-\Theta(c|x)\right]\Pc\Big].
    \label{eq:evol-PC}
\end{align}
This is not a closed equation, $C$ alone is not enough to characterise
the stochastic process. The delta functions express the fact that the
marginal distribution only changes, logically, at the control updates.

In the main text, we have introduced the notation $P^{(k)}(c)\equiv P(c,t_{k}^{+})$, which gives the probability of the control values throughout the whole time interval $I_{k}$. The constancy of $P^{(k)}(c)$ over the time interval $I_k$ entails that the conditional probability $P(x,t|c)$ obeys the same Fokker-Planck equation as $\Pc$ between control updates. Therefore, $P(x,t|c)$ irreversible evolves towards the stationary solution $\PsXC$ of the Fokker-Planck equation corresponding to the potential $\tilde{V}(x,c)$, which verifies
\begin{equation}\label{eq:FP-Jk-v2}
    \tilde{V}'(x,c) \PsXC+ k_{B}T\,\partial_{x}
    \PsXC=-\gamma J_{\sts}^{(c)} ,
\end{equation}
with $J_{\sts}^{(c)}$ being the uniform steady current with the control variable equal to $c$.  { Note that, in general, $P(x,t|c)$ does not reach $\PsXC$ at the end of the interval $I_{k}$. At that instant, the Fokker-Planck evolution is interrupted by the action of the control, which changes discontinuously the value of $\Pc$. It is only in the limit as $\Delta t_{m}\to\infty$ that $P(x,t|c)$ reaches $\PsXC$---see
Sec.~\ref{sec:delta-tm-infty}.}

For later reference, we write here
the expression for $\PsXC$ and $J_{\sts}^{(c)}$ for the periodic boundary
conditions in Eq.~\eqref{eq:periodic-bc}, which are basically the transposition to the case considered here of the formulas just derived above:\begin{equation}\label{eq:Psts_x|c}
  \PsXC=K(c) \left[e^{-\beta \tilde{V}(x,c)}- \frac{e^{-\beta \tilde{V}(´L,c)}-e^{-\beta
        \tilde{V}(-L,c)}}{\int_{-L}^{L}dx'
  e^{-\beta\left[\tilde{V}(L,c)-\tilde{V}(x',c)\right] }  }\int_{-L}^{x}dx'
  e^{-\beta\left[\tilde{V}(x,c)-\tilde{V}(x',c)\right] }  \right] ,
\end{equation}
where $K(c)>0$ is a normalisation constant, such that
$\int_{-L}^{L}dx\, \PsXC=1$. The corresponding steady current is
\begin{equation}
  \beta\gamma J_{\sts}^{(c)}=K(c) \frac{e^{-\beta \tilde{V}(L,c)}- e^{-\beta \tilde{V}(-L,c)}}
  {\int_{-L}^{L}dx\,
  e^{-\beta\left[\tilde{V}(L,c)-\tilde{V}(x,c)\right] } }.
\end{equation}
The average velocity of the particle in the steady
state corresponding to $\tilde{V}(x,c)$ is $\av{v}_{\sts}(c)=2LJ_{\sts}^{(c)}$. In agreement with physical
intuition, the particle moves to the right in the steady state for
$\tilde{V}(x,c)$, i.e. $\av{v}_{\sts}(c)>0$ (or $J_{\sts}^{(c)}>0$), when
$\tilde{V}(L,c)<\tilde{V}(-L,c)$.\footnote{In the absence of external force,
$F_{\ext}=0$, the periodicity of the potential $V(x,c)$ ensures that $J_{\sts}^{(c)}=0$ and
the stationary distributions have the canonical form,
$\PsXC=P^{\eq}(x|c)\propto \exp[-\beta V(x,c)]$.}

\section{Feedback flashing protocol}\label{app:f-f-protocol}

Here, we explicitly write the expression for the flashing protocol presented in Fig.~\ref{fig:protocols} and studied in Sec.~\ref{sec:flashing-ratchet}, with uncertainty $\Delta x$ in the measurement of the position. In order to obtain an easier expression in terms of the modulo function $\modulo(x,L)$, we shift the spatial interval by  $+L$, i.e. we consider periodic boundary conditions in $[0,2L]$. With this choice, for a generic saw-tooth potential $V(x)$ with minimum at $x=a\in (0,2L)$, we have
\begin{equation}
  \Theta(1|x) =
\begin{cases}
    1, & \text{if } \lb <\ub\leq a, \\
    \frac{a-\lb}{2 \Delta x}, & \text{if } \lb \leq a < \ub, \\
    0, & \text{if } a < \lb \leq \ub, \\
    \frac{a+\ub-\lb}{2 \Delta x}, & \text{if } \ub \leq a < \lb, \\
    \frac{\ub}{2 \Delta x}, & \text{if } \ub \leq a < \lb, \\
    \frac{a}{2 \Delta x}, & \text{if } a < \ub \leq \lb .
\end{cases}  
\end{equation}
For the symmetric V-shaped potential considered in our numerical simulations, $a=L$. Once we have specified $\Theta(1|x)$, it is straightforward to obtain $\Theta(0|x)=1-\Theta(1|x)$.

\section{Eigenvalues and eigenfunctions of the Fokker-Planck equation with the V-shaped potential} \label{app:eigenvalues-and-eigenfunctions}

Let us consider the Green function $K(x,t|x',c)$ for the Fokker-Planck equation 
particularised for the V-shaped potential defined in Eq. \eqref{eq_V-shape}. Then, we have to solve 
\begin{align} \label{eq:fp}
    \gamma\,\partial_t K(x,t|x',c)=& \partial_x \left[c\, F \sgn(x)\,K(x,t|x',c)\right]+\beta^{-1} \partial^2_x K(x,t|x',c),
\end{align}
with  initial condition
\begin{equation}
   K(x,t=0|x',c)=\delta(x-x'),
\end{equation}
and periodic boundary conditions.

The Green function is given in general by Eq.~\eqref{eq:green-function} of the main text in terms of the the eigenvalues and eigenfunctions of the Fokker-Planck equation with operator $\calL^{(c)}$, i.e. for the potential $V(x,c)$. The stationary distributions $\PsXC$, corresponding to the non-degenerate eigenvalue $\lambda_0^{(c)}=0$,  have been derived in \ref{app:cond-steady-distr} for a general potential $V(x,c)$. Particularising these expressions for our V-shaped potential $V(x,c)= c\, F \abs{x}$, Eq.~\eqref{eq:marginal-steady-state} of the main text is obtained. The remainder of eigenvalues $\lambda_n^{(c)}>0$ are doubly degenerate and, therefore, the Green function can be written, more specifically, as
\begin{align}
    K(x,c|x',t)=\PsXC+
    \frac{1}{\PsXprimeC}\sum_{n=1}^{\infty} e^{-\lambda^{(c)}_n t}\Big[ \phi^{(c)}_{n,1}(x')\phi^{(c)}_{n,1}(x)+\phi^{(c)}_{n,2}(x')\phi^{(c)}_{n,2}(x)\Big]
\end{align}
with eigenvalues
\begin{align}
   & \lambda_0^{(1)}=0,\quad \lambda_n^{(1)}= \frac{\beta F^2}{4} + \frac{n^2 \pi^2}{\beta L^2}, && n=1,\dots,\\  &\lambda_n^{(0)}=\frac{n^2 \pi^2}{\beta L^2}, && n=0,1,\dots,
\end{align}
and eigenfunctions
\begin{subequations}
  \begin{align}
    \phi^{(1)}_0(x)&=\frac{F \beta}{2(1-e^{-F \beta L}) } e^{-F\beta \abs{x}}, &&
    \phi^{(0)}_0(x)= \frac{1}{2L}, \\
   \phi^{(1)}_{n,1}(x)&=\frac{\sqrt{\phi^{(1)}_0(x)}}{\sqrt{L}}\sin\left(\frac{n\pi x}{L}\right), &&
   \phi^{(0)}_{n,1}(x)= \frac{1}{\sqrt{2}L} \cos\left(\frac{n \pi x}{L}\right), \\
  \phi^{(1)}_{n,2}(x)&=\frac{2\sqrt{\phi^{(1)}_0(x)}}{\sqrt{L\beta \lambda^{(1)}_n}}\left[\frac{2 n \pi}{L} \cos\left(\frac{n \pi x}{L}\right)-F \beta \sin\left(\frac{n\pi \abs{x}}{L}\right)\right], &&
   \phi^{(0)}_{n,2}(x)= \frac{1}{\sqrt{2}L} \sin\left(\frac{n \pi x}{L}\right).
\end{align}  
\end{subequations}

\section{Calculation of the coefficients $B_{cc'}$}\label{app:computation-of-bcc}

In Sec.~\ref{sec:flashing-ratchet}, we have computed the coefficients $B_{cc'}$ defined in Eq.~\eqref{eq:matrixB-def}, for the particular protocol we have considered, with error $\Delta x$. All the coefficients $B_{cc'}$ are equal to $1/2$ in that case. Interestingly, this result is more general and applies to symmetric potentials and flashing protocols with the following structure:
\begin{align}\label{eq:theta1-sym}
\Theta(1|x)=f_1(x)\,\eta(x)+(1-f_1(-x))\,\eta(-x),
\end{align}
where $\eta(x)$ represents the Heaviside function, and $0\leq f_1(x)\leq 1$.\footnote{The latter condition ensures that $0\le \Theta(1|x)\le 1$.} Of course, $\Theta(0|x)=1-\Theta(1|x)=(1-f_1(x))\,\eta(x)+f_1(-x)\,\eta(-x)$.
Let us calculate $B_{1c}$, assuming that $\Theta(1|x)$ has the structure in Eq.~\eqref{eq:theta1-sym}:
\begin{align}
    B_{1c}&=\int dx \, \Theta(1|x)\PsXC=\int_{0}^L dx \, f_1(x) \PsXC +\int_{-L}^0 dx \, (1-f_1(-x))\PsXC \nonumber \\
    &=\int_{0}^L dx \, f_1(x) \PsXC +\int_{0}^L dx \, (1-f_1(x))\PsXC=\int_{0}^L dx \,  \PsXC=\frac{1}{2},
\end{align}
where we have employed that $\PsXC=P_{\sts}^{(c)}(-x)$, due to the symmetry of the potential. An analogous calculation shows that $B_{0c}=1/2$, since $\Theta(0|x)=1-\Theta(1|x)$.

The class of flashing protocols in Eq.~\eqref{eq:theta1-sym} has a neat physical interpretation. For the V-shaped potential, we recall that the error-free case corresponds to activating the potential for $x<0$. Therefore, Eq.~\eqref{eq:theta1-sym} tells us that the probability of wrongly switching on the potential is $\Theta(1|x>0)=f_1(x)$ for $x>0$.  Also, it tells us that the probability of correctly switching on the potential is $\Theta(1|x<0)=1-f_1(\abs{x})$. Therefore, the probability of success---switching on the potential correctly---and the probability of error---switching on the potential wrongly---are balanced, since both of them add to one. Note that this condition is not necessary for $\Theta(c|x)$ to be a well-defined probability, but it seems to be a reasonable requirement at the physical level that is fulfilled  by the protocol  we have analysed in this paper. In Fig.~\ref{fig:protocols}, this condition is clearly illustrated, since the protocol is odd with respect to the $y=1/2$ axis. 

Finally, we would like to remark that the structure \eqref{eq:theta1-sym} is also present in other ``noisy" flashing protocols considered in the literature. For example, the flashing protocol of Ref.~\cite{cao_information_2009} is written in our notation as
\begin{align}\label{eq:cao-touchette-protocol}
    \Theta(1|x)=p \,\eta(x)+(1-p)\,\eta(-x), \quad 
    {\color{black}\Theta(0|x)=1-\Theta(1|x)=(1-p) \,\eta(x)+p\,\eta(-x),}
\end{align}
i.e. $f_1(x)=p$, independent of $x$. Therefore, the error in this flashing protocol is simpler than the one considered in this paper: the probability of wrongly switching on the potential  is equal to $p$, regardless of the position of the particle. Note that in our flashing protocol with error in the measurement $\Delta x$, the probability of wrongly switching on the potential increases as the particle approaches the point $x=0$ at which the force changes sign. 

{\color{black} The mutual information between the particle's position
  and the control state follows from our Eq.~\eqref{eq:mutual-information-all-tm}:
\begin{align}
    I_{\sts}(X,C) =&-\sum_c P_{\sts}(c)\ln{P_{\sts}(c)} +\int  dx \, P_{\sts}(x) \sum_c \Theta(c|x) \ln{\Theta(c|x)}\nonumber \\
    =& -\sum_c P_{\sts}(c)\ln{P_{\sts}(c)} +\int_{0}^{L}  dx \, P_{\sts}(x) \,  p \ln{p} +\int_{-L}^{0}  dx \, P_{\sts}(x)\,  (1-p) \ln{(1-p)} \nonumber \\
    & + \int_{0}^{L}  dx \, P_{\sts}(x) \,  (1-p) \ln(1-p) +\int_{-L}^{0}  dx \, P_{\sts}(x) \, p \ln{p}\nonumber \nonumber\\
      =& -\sum_c P_{\sts}(c)\ln{P_{\sts}(c)} +p \ln{p} + (1-p) \ln{(1-p)}. \label{eq:Hq-Hp}
    \end{align}
This equation is equivalent to Eq.~(6) in
Ref.~\cite{cao_information_2009}. This can be made more transparent by
introducing the notation employed therein, (i) for the entropy of a dichotomous variable with probability $p$ as
\begin{equation}
    H(p) \equiv -p \ln p - (1-p) \ln (1-p),
\end{equation}
and (ii) for the probability of switching off the potential as
\begin{equation}
    q \equiv P_{\sts}(c=0) =(1-p)\int_{0}^L \, dx \, P_{\sts}(x) + p  \int_{-L}^0\, dx \,P_{\sts}(x),
\end{equation}
which follows by combining Eqs.~\eqref{eq:P-k-C-sts} and \eqref{eq:cao-touchette-protocol}. Of course,
\begin{equation}
    1-q=P_{\sts}(c=1)=p\int_{0}^{L} \,dx \, P_{\sts}(x)+(1-p)\int_{-L}^{0} \, dx \, P_{\sts}(x).
\end{equation}
With these definitions, 
\begin{equation}
    I_{\sts}(X,C)=H(q)-H(p),
\end{equation}
which is precisely Eq.~(6) of Ref.~\cite{cao_information_2009}, which can thus be regarded as a particular case of the general expression~\eqref{eq:mutual-information-all-tm} for the mutual information derived in the wide framework presented in our paper.
}

\section*{References}
\bibliographystyle{iopart-num}
\bibliography{Mi-biblioteca-30-may-2024}

\end{document}